\documentclass[preprint,aps,tightenlines,floatfix,showpacs]{revtex4}

\usepackage{graphicx}
\usepackage{bm}
\usepackage{amsmath}
\usepackage{amssymb}

\newcommand{\sfrac}[2]{\textstyle\frac{#1}{#2}}

\begin{document}

\title{A multichannel model for clusters of an $\alpha$ and select $N=Z$
nuclei}

  \author{K. Amos$^{(1,4)}$}
  \email{amos@unimelb.edu.au}
  \author{L. Canton$^{(2)}$}
  \author{P. R. Fraser$^{(1,3)}$}
  \author{\mbox{S. Karataglidis$^{(1,4)}$}}
  \author{J. P. Svenne$^{(5)}$}
  \author{D. van der Knijff$^{(1)}$}

  \affiliation{$^{(1)}$ School  of Physics,  University of  Melbourne,
    Victoria 3010, Australia}

  \affiliation{$^{(2)}$ Istituto  Nazionale  di  Fisica  Nucleare,
    Sezione di Padova, Padova I-35131, Italia}

  \affiliation{$^{(3)}$ 
ARC Centre for Antimatter-Matter Studies, Curtin University, GPO Box U1987,
Perth, Western Australia 6845, Australia}

  \affiliation{$^{(4)}$ Department of Physics, University of Johannesburg,
    P.O. Box 524 Auckland Park, 2006, South Africa}

  \affiliation{$^{(5)}$ Department  of  Physics  and Astronomy,
    University of Manitoba, and Winnipeg Institute for Theoretical Physics,
    Winnipeg, Manitoba, Canada R3T 2N2}

\date{\today}

\begin{abstract}
A multi-channel algebraic scattering (MCAS) method has been used 
to solve coupled sets of Lippmann-Schwinger equations for $\alpha$+nucleus
systems to find spectra of the compound systems. 
Low energy spectra for ${}^{12}$C, ${}^{16}$O, and ${}^{20}$Ne are found 
with the systems considered as the coupling of an $\alpha$ particle with 
low-excitation states of the core nuclei, ${}^8$Be, ${}^{12}$C, and
${}^{16}$O, respectively.  Collective models have been used to define
the matrices of interacting potentials. Quadrupole (and octupole when
relevant) deformation is allowed and taken to second order.  
The calculations also require a small monopole interaction to provide an 
extra energy gap commensurate with an effect of strong pairing forces.
The results compare reasonably well with known spectra given
the simple collective model prescriptions taken for the coupled-channel
interactions. Improvement of those interaction specifics in the approach
 will give spectra and wave functions suitable for use in analyses of 
cross sections for $\alpha$ scattering and capture by light-mass nuclei;
reactions of great importance in nuclear astrophysics.
\end{abstract}

\maketitle

\section{Introduction}

In light stars ($\le$ 1.5 M$_\odot$), proton-proton chain reactions
lead to the formation of nuclei up to mass-8. Once the $\alpha$ particles
generated in those reactions are present
in sufficient number, the triple-$\alpha$ process can produce ${}^{12}$C; the crucial
feature being the energy of the Hoyle state in ${}^{12}$C lying
just above the break-up threshold.
Thereafter, $\alpha$-particle captures play important roles in the catalytic C-N-O  
cycle of nucleogenesis. Thus the formation of light-mass nuclei
by $\alpha$ capture at low energies, and their spectra, especially of the
resonances lying above the $\alpha$-breakup thresholds, are of much 
interest~\cite{Th09}.

Treating nuclei as a cluster of two or more composite particles
has a long history. Notably there are states of some light-mass nuclei 
that can be interpreted as clusters of $\alpha$ particles, 
sometimes with additional valence nucleons.  The Hoyle state in 
${}^{12}$C is perhaps the most famous so treated.  Clustering has been 
found influential particularly for states in nuclei that lie close to 
relevant decay thresholds~\cite{Oe06}.  Cluster model approaches 
also have been applied to assess the properties of the ground and 
sub-threshold states in nuclei; properties which often are also well 
described by many-nucleon models of structure. 
Likewise many cluster-model theories of nuclear structure and reactions 
have been developed with Ref.~\cite{Fr07} being a comprehensive review.
As well, there is a review~\cite{My14} of the use of a
complex scaling method with a cluster orbital shell model 
to find many-body resonances and the continua in light nuclei.

Recently, several systems of an $\alpha$ and an (even-even) nucleus were treated 
theoretically~\cite{Ye12b} with a semi-algebraic cluster model~\cite{Ye12,Fr12}, which
accounts for Pauli-blocking of the constituent nucleons of clusters in the compound
system, but only states explicitly known to be populated by these clusterisations were
studied.  Two of those systems, $\alpha+{}^{12}$C and $\alpha+{}^{16}$O, are considered 
this work.  Even more recently, the low-excitation, positive-parity states of ${}^{16}$O
were well described by $\alpha+{}^{12}$C clusters using the generator
coordinate method with wave functions given by the antisymmetrized
molecular dynamics scheme~\cite{KE14}. That article~\cite{KE14}
contains an extensive set of references relating to such cluster model studies.

 Herein, we consider the low-excitation spectra of ${}^{12}$C, ${}^{16}$O, and 
 of ${}^{20}$Ne by using a multi-channel algebraic scattering (MCAS) method~\cite{Am03}
 for the compound system of an $\alpha$-particle interacting with low-excitation collective 
 states of the core nuclei, ${}^8$Be, ${}^{12}$C, and ${}^{16}$O,
 respectively.  
 Experimental values for the states of ${}^8$Be were taken from Ref.~\cite{Ti04},
those for ${}^{12}$C from Ref.~\cite{Aj90}, those for ${}^{16}$O from
Ref.~\cite{Ti93}, and those for ${}^{20}$Ne from Ref.~\cite{Ti98}.
 With the current form of MCAS,  descriptions of the isoscalar
 states in the compound systems are found.  Each case in the set considered has some
 uniquely problematic aspects.  In this, our first study of these systems using MCAS, 
 we consider mainly the spectra of the compound nuclei.  
 The $\alpha$-emission thresholds of the three nuclei, ${}^{12}$C, ${}^{16}$O, and
 ${}^{20}$Ne, lie between 7 and 8 MeV excitation, only a few MeV above which,  the
 density of levels in the compound systems increases rapidly. 
However, MCAS is able to determine elastic 
scattering cross sections and we present herein also a first test
estimate for an $\alpha+{}^{16}$O cross section.

We take  ${}^8$Be as the core for the cluster view of ${}^{12}$C.
The ground state of $^8$Be lies just above the $\alpha$-particle 
breakup threshold  and its two lowest excited 
states are broad resonances. They have spin-parities $2^+$ and $4^+$
with  energy centroids (widths) of 3.03 (1.51) MeV and of 11.35 
(3.50) MeV respectively. 
These three resonance states of ${}^8$Be
have been considered in the coupled-channel calculation of ${}^{12}$C taken
as the $\alpha+{}^8$Be cluster. Their resonance
properties are known to have an impact in cluster model evaluations of the 
spectra of the compound nuclei~\cite{Fr08,Ca11}.

Four low-lying states in ${}^{12}$C, the ground ($0^+$), 
the $2_1^+$ (4.44 MeV), the $0_2^+$ (7.65 MeV), and the $3^-$ (9.64 MeV)
have been used in our MCAS evaluations of the spectrum of the compound
$^{16}$O. The ground and $2^+_1$ states are stable against particle
emission and the two higher lying ones can decay by $\alpha$ emission
but their widths are sufficiently small that there is only minor effects
due to those aspects.

In the two evaluations mentioned above, the interaction potentials have 
been formed assuming a 
collective rotational model having isoscalar quadrupole and, with 
negative-parity target states,
octupole deformation. An intriguing question is to find how that can
lead to a low-excitation spectrum (for ${}^{16}$O) that is usually
considered as a set of vibrations upon a closed-core ground state.

Finally in treating ${}^{20}$Ne as an $\alpha+{}^{16}$O system,
besides the ground state, we have used the $0_2^+$ (6.05 MeV), the
$3^-$ (6.13 MeV), and the $2^+$ (6.92 MeV) states in ${}^{16}$O in forming the
coupled-channel Hamiltonian.  In this case, we have generated the interaction
potentials by using a vibration model for the $\alpha+{}^{16}$O
cluster and an intriguing question posed in this case is to find 
the more rotor-like low-energy spectrum of ${}^{20}$Ne.
In these pursuits, of course, we presume that all low-excitation
isoscalar states in the compound systems, whatever their exact
description, will have components that overlap with the cluster
of an $\alpha$ and core nuclei.  We presume also that only strongly
coupled collective states in those core nuclei contribute importantly
in defining the Hamiltonians.

A pr\'ecis  of the MCAS approach is given next.
Then, in Sec.~\ref{potmat}, specifics of the matrix of potentials
for an $\alpha$+nucleus system are defined. 
As a first test of the $\alpha+A$ MCAS code,  
it was used to find spectra for  ${}^7$Li ($\alpha+{}^3$H) and ${}^7$Be
($\alpha+{}^3$He); spectra found previously~\cite{Ca06a} but using a version
of the code built for incident spin-$\frac{1}{2}$ particles.
Test results are discussed in Sec.~\ref{test}.
In Sec.~\ref{carbon}, we show the results for the spectrum
of ${}^{12}$C considered as the $\alpha+{}^8$Be cluster. 
Sec.~\ref{oxygen} contains the MCAS results for the spectrum
of ${}^{16}$O treated as an $\alpha+{}^{12}$C system and 
the MCAS results for ${}^{20}$Ne (as $\alpha+{}^{16}$O) 
are given in Sec.~\ref{neon}. 
Finally, in Sec.~\ref{elscat}, MCAS results for an elastic scattering cross section
 of $\alpha$ particles from ${}^{16}$O are shown to illustrate
the utility that the approach may have in scattering data analysis.
Conclusions are made in Sec.~\ref{conclusion}.

\section{A pr\'ecis of MCAS}
\label{precis}

The MCAS approach is a method to solve coupled Lippmann-Schwinger (LS) 
equations for a chosen two-cluster interaction matrix of potentials.
The method uses  separable expansions of those matrices of 
potentials, where a crucial choice is that of the expansion basis.
The optimal choices of the form factors of those separable expansions 
are those derived from sturmian functions determined from the 
specifically-chosen two-cluster interaction matrix of potentials.

Consider a system of $\Gamma$ channels for each allowed scattering 
spin-parity, $J^\pi$, with the index $c\ (=1,\Gamma)$ denoting the set of 
quantum numbers that identify each channel uniquely. 
The integral equation approach in momentum space for potential matrices, 
$V_{cc'}^{J^\pi}(p,q)$, requires solution of coupled LS 
equations giving a multichannel $T$ matrix of the form
\begin{align}
T_{cc'}^{J^\pi}(p,q;E)\ =\ V_{cc'}^{J^\pi}(p,q) + \mu& \left[ 
\sum_{c'' = 1}^{\rm open}
\int_0^\infty V_{cc''}^{J^\pi}(p,x) \frac{x^2}{k^2_{c''} - x^2 + i\epsilon}
T_{c''c'}^{J^\pi}(x,q;E)\ dx \right. 
\nonumber\\
&\;\; \left.- \sum_{c'' = 1}^{\rm closed} \int_0^\infty
V_{cc''}^{J^\pi}(p,x) \frac{x^2}{h^2_{c''} + x^2} 
T_{c''c'}^{J^\pi}(x,q;E) \ dx \right] .
\label{multiTeq}
\end{align}
Therein the contributions from open and closed channels have been separated 
with the respective channel wave numbers being 
\begin{equation}
k_c = \sqrt{\mu(E - \epsilon_c)}\hspace*{1.0cm} h_c = 
\sqrt{\mu(\epsilon_c - E)}\ ,
\label{wavenos}
\end{equation}
for $E > \epsilon_c$ and $E < \epsilon_c$ respectively with $\epsilon_c$ 
being the threshold energy of channel $c$. Here $\mu$ designates 
$2m_{\rm red}/\hbar^2$ with $m_{\rm red}$ being the reduced mass.  With the  
${J^\pi}$ superscript understood from now on, solutions of 
Eq.~(\ref{multiTeq}) are sought using expansions of the potential matrix 
elements in (finite) sums of energy-independent separable terms,
\begin{equation}
V_{cc'}(p,q) \sim  \sum^N_{n = 1} {\hat \chi}_{cn}(p)\
\eta^{-1}_n\ {\hat \chi}_{c'n}(q)\ .
\label{finiteS}
\end{equation}
With these expansions the multichannel $S$-matrix acquires, in general,
 a closed algebraic form.
Indeed, the link between the multichannel $T$ matrix and the scattering matrix
is~\cite{Ca91,Pi95}
\begin{align}
S_{cc'}\ =&\ \delta_{cc'}\ -\ i \pi \mu \sqrt{k_c k_{c'}}\ T_{cc'}
\nonumber\\
=&\ \delta_{cc'}\ -\ i^{\left(l_{c'} - l_c +1\right)} 
\pi \mu \sum_{n,n' = 1}^N \sqrt{k_c}\ 
{\hat \chi}_{cn}(k_c)\ \left([\mbox{\boldmath $\eta$} - {\bf G}_0]^{-1}
\right)_{nn'}\ {\hat \chi}_{c'n'}(k_{c'})\sqrt{k_{c'}}\ ,
\label{multiS}
\end{align}
where now $c,c'$ refer to open channels only.  In this representation, and in
the case of discrete target states, \textbf{${{\bf G}_0}$} 
and \mbox{\boldmath $\eta$} have matrix elements (for each
value of $J^\pi$ being understood)
\begin{align}
\left[{\bf{G}}_0 \right]_{nn'} =&\ \mu\left[ \sum_{c = 1}^{\rm open} \int_0^\infty
{\hat \chi}_{cn}(x) \frac{x^2}{k_c^2 - x^2 + i\epsilon} {\hat 
\chi}_{cn'}(x)\ dx -
\sum_{c = 1}^{\rm closed} \int_0^\infty {\hat \chi}_{cn}(x) 
\frac{x^2}{h_c^2 + x^2}
{\hat \chi}_{cn'}(x)\ dx \right]
\nonumber\\
\left[{\mbox {\boldmath $\eta$ }}\right]_{nn'} =&\ \eta_n\ \delta_{nn'} \, .
\label{xiGels}
\end{align}
The bound states of the compound system are defined by the zeros of the matrix
determinant when the energy is $E < 0$ and so link to the zeros of
$\{ \left| \mbox{\boldmath $\eta$}-{\bf G}_0\right| \}$ when all channels in 
Eq.~(\ref{xiGels}) are closed.

With coupling involving bound target states, the usual method of
solution of the LS coupled equations uses the method of principal parts.
However, 
when the target states inherent in the system are themselves resonances,
the propagators in the LS coupled equations must be suitably
modified~\cite{Ca11} and direct evaluation of the
integrals having complex kernels becomes possible.

\section{The model for the $\alpha$+nucleus matrix of potentials}
\label{potmat}

We find the $\alpha$+nucleus matrix of potentials by  using a 
collective model for the structures of the target nuclei.  
All terms, to second order 
in deformation, are carried given that the collectivity of the nucleus 
studied may be strong. Also, the potential field is allowed to  have central 
($V_0$), $\ell^2$-dependent ($V_{\ell \ell}$), target state spin-dependent
($V_{II}$), 
and orbit-nuclear spin ($V_{\ell I}$) components.  
An extra monopole interaction is allowed in the interaction
between the $\alpha$ and the target in its ground ($0^+$) state.

Consider a basis of channel states defined by the coupling
\begin{equation}
\left| c \right\rangle = \left| \ell I J^\pi \right\rangle =
\biggl[
\left| \ell\right\rangle \otimes \left| \psi_I \right\rangle
\biggr]_J^{M,\pi}\ ,
\label{ch-state}
\end{equation}
where $\ell$ is the orbital angular momentum of relative motion of 
a spin-0 projectile on the target whose  
states are $\left| \psi_I^{(N)} \right\rangle$.
With each $J^\pi$ hereafter understood, and by disregarding deformation 
temporarily, the ($\alpha$+nucleus) potential matrices may be written,
as
\begin{align}
V_{cc'}(r) = \langle \ell I  \left|\ W(r)\ \right|
\ell' I' \rangle
=& \bigg[ V_0 \delta_{c'c} f(r) + V_{\ell \ell} f(r) 
[ {\bf {\ell \cdot \ell}} ]  + V_{II} f(r) [{\bf I \cdot I}] 
+  V_{\ell I} g(r) [ {\bf {\ell \cdot I}} ]\bigg]_{cc'}
\nonumber\\
&\hspace*{0.5cm} + V_{mono} \delta_{c'c} \delta_{I0^+_{g.s}} f(r),
\label{www1}
\end{align}
in which local form factors have been assumed.  Typically they are specified as
Woods-Saxon functions,
\begin{equation}
f(r) = \left[1 + e^{\left( \frac{r-R}{a} \right)} \right]^{-1}
\hspace*{0.3cm} ; \hspace*{0.3cm} g(r) = \frac{1}{r} \frac{df(r)}{dr} .
\label{radforms}
\end{equation}
A monopole term is included to allow $N=Z$ even-mass systems 
to have states in which pairing effects lead to extra binding.

Deformation then is included with the nuclear surface defined by 
$R(\theta,\phi) = R_0 ( 1 + \epsilon)$ wherein $\epsilon$ is a generic term
to be specified according to whether a rotational or a vibrational
collective model for nuclei is used. Details are given in the appendices.
Treating $R(\theta \phi)$ as the variable in $f(r) = f(r-R(\theta \phi))$,
the function, $f(r)$, on expanding in the deformation to order $\epsilon^2$,
becomes
\begin{equation}
f(r) = f_0(r) + \epsilon \left[\frac {df(r)}{d \epsilon}\right]_0
+ \frac {1}{2} \epsilon^2 \left[ \frac {d^2 f(r)}{d \epsilon^2}\right]_0
= f_0(r) - R_0 \frac{df_0(r)}{dr}\ \epsilon
+ \frac {1}{2} R_0^2\  \frac {d^2 f_0(r)}{d r^2}\ \epsilon^2 .
\label{Eqn4}
\end{equation}
The subscript `0' indicates the spherical Woods-Saxon form 
with $R = R_0$. 
There is a similar equation for $g(r)$. 

More details of the expansion of these matrix elements 
for an $\alpha$+nucleus cluster are given
in Appendix~\ref{app1}.   Specifics of these potential elements when a
rotation model of collectivity is used are given in Appendix~\ref{app2}
while the details relevant to use of a vibration model  are given
in Appendix~\ref{app3}.

When collective models are used to specify the matrix of
interaction potentials acting between a nuclear projectile and the 
target nucleus in which a set of states are active, there are problems in satisfying 
the Pauli principle~\cite{Ca05}.  In the MCAS method the effects of the 
Pauli principle are met by inclusion of a set of orthogonalizing 
pseudo-potentials (OPP)~\cite{Am03}; a technique that was developed in 
studies of cluster physics~\cite{Kr74,Ku78} as a variant of the Orthogonality
Condition Model (OCM) of Saito~\cite{Sa69} and, more recently,
in a study~\cite{Ho13} using few-body models to specify cluster
structure in light nuclei.
In this way, the effects of 
Pauli blocking in the relative motion of two clusters comprised of fermion constituents
could be taken into account.
The OPP can also be used for the situation with partially occupied
levels being Pauli hindered. Schmid~\cite{Sc78} notes that states can be
Pauli-forbidden, Pauli- allowed, or Pauli-suppressed; the last being what 
we have called Pauli hindrance in MCAS theory~\cite{Ca06,Am12,Am13}. 

To orthogonalize states describing intra-cluster motion with respect to
the deeply-bound Pauli forbidden states, MCAS uses highly nonlocal OPP
terms embedded in a coupled-channel context. 
The matrix of interaction potentials (in coordinate space) to be used
has the form
\begin{equation}
{\cal V}_{cc'} = V_{cc'}(r) \delta(r-r') + 
\lambda_c A_c(r) A_{c'}(r') \delta_{cc'} .
\end{equation}
$V_{cc'}(r)$ is the nuclear interaction potential and
$\lambda_c$ is the scale we use to give Pauli blocking. 
The $A_{c}(r)$ are bound-state wave functions of the $\alpha$ in the diagonal
potentials, $V_{cc}(r)$, for each target state in channel $c$.
Pauli blocking of the specific orbit in a particular channel, $c$, is
achieved by using a very large value for $\lambda_c$. That value should
be infinite but for all practical purposes $10^6$ MeV suffices. 
Pauli allowed states have $\lambda_c = 0$ while Pauli hindrance is achieved 
by using intermediate values for $\lambda_c$.


\section{Test cases: MCAS and  
$\alpha+^3{\rm H}$, $\alpha+^3{\rm He}$, and $\alpha+\alpha$}
\label{test}

These test cases are taken to be single channel problems given that
the components are quite strongly bound and have no excited states 
below nucleon emission thresholds. But the compound systems do have 
well established spectra and, for the $\alpha+^3$H and $\alpha+^3$H${\rm e}$
systems, the states that we might expect to obtain are those 
indicated  in Table~\ref{7Li-states}. The reactions involving an $\alpha$ 
that lead to them, or have the mass-7 states as a compound system, are indicated
by the check marks.
\begin{table}[h]
\caption{\label{7Li-states} 
  States in  ${}^{7}$Li and of ${}^7$Be relevant to this investigation
  and known reactions~\cite{Fi96}  involving an $\alpha$ that populate them.}
\begin{ruledtabular}
\begin{tabular}{l|ccc|ccc}
J$^\pi$ & & ${}^7$Li & & & ${}^7$Be & \\
\hline
$J^\pi$ & $^3$H$(\alpha,$n$)$ & $^4$He$(^3$He$,\pi^+)$ & $^4$He$(\alpha,$p$)$
& $^4$He$(^3$He$,\gamma)$ & $^4$He$(^3$He$,^3$He$),(^3$He$,$p$)$ & $^4$He$(\alpha,$n$)$\\
\hline
$\sfrac{3}{2}^-$ & $\surd$  & $\surd$ & $\surd$ 
& $\surd$ &   & $\surd$ \\
$\sfrac{1}{2}^-$ &   & $\surd$ & $\surd$ 
& $\surd$ &   & $\surd$ \\
$\sfrac{7}{2}^-$ &   &   & $\surd$ 
&   & $\surd$ & \\
$\sfrac{5}{2}^-$ &   & $\surd$ &   
&   & $\surd$ & \\
\hline
\end{tabular}
\end{ruledtabular}
\end{table}

\subsection{The $\alpha+^3${\rm H} and $\alpha+^3${\rm He} systems}

Spectra of ${}^7$Li and ${}^7$Be have been found previously~\cite{Ca06a} 
using the MCAS program written for spin-$\frac{1}{2}$ particles coupling
to a nucleus, i.e. as ${}^3$H+$\alpha$ and ${}^3$He+$\alpha$ systems
respectively. The results agreed well with known states in those
spectra. Thus the first study made with the MCAS program written for
$\alpha$ (spin-0) particles coupling to a nucleus has been of these
systems but taken as an $\alpha$-particle coupling to the two 
mass-3 systems as the core nuclei. 
\begin{table}[h]
\caption{\label{mass7}
Spectra of ${}^7$Li and ${}^7$Be from $\alpha$
coupled to ${}^3$H and ${}^3$He respectively.
The energies are in MeV while the widths are in keV.
The experimental values are those listed in~\cite{Ti02}.
The results labelled `previous'  are from~\cite{Ca06a}}
\begin{ruledtabular}
\begin{tabular}{c|ccc|ccc}
 & & ${}^7$Li & & & ${}^7$Be & \\
$J^\pi$ & Exp. & present & previous &
 Exp. & present & previous \\
\hline
$\frac{3}{2}^-$ & spurious  & $-$29.6 & $-$29.4  
& spurious & $-$27.8 & $-$28.0 \\
$\frac{1}{2}^-$ & spurious  & $-$28.0 & $-$27.8
& spurious & $-$26.3 & $-$26.4 \\
\hline
$\frac{3}{2}^-$ & $-$2.47   & $-$2.59 & $-$2.47
& $-$1.59 & $-$1.53 & $-$1.53\\
$\frac{1}{2}^-$ & $-$1.99   & $-$1.87 & $-$1.75
& $-$1.16 & $-$0.85 & $-$0.84 \\
$\frac{7}{2}^-$ & 2.18\ (69) & 2.09\ (80) & 2.12 (83)
& 2.98\ (175$\pm$ 7)& 3.14\ (204) & 3.07 (180)\\
$\frac{5}{2}^-$ & 4.13\ (918) & 4.05\ (800) & 4.12 (834)
& 5.14\ (1200)& 5.13\ (1250) & 5.09 (1194)\\
\end{tabular}
\end{ruledtabular}
\end{table}
The interaction form specified in Eq.~(\ref{www1}) was used 
with strength parameter values (in MeV),
$V_0$ = -76.8, $V_{l l}$ = 1.15, and  $V_{l I}$ = 2.34.
The geometry of the Woods-Saxon form was set with 
$R_0$ = 2.39 and $a$ = 0.68 fm. The Coulomb potential was that from
a uniformly charged sphere of radius $R_c$ = 2.34 fm.  
A slightly larger charge radius (2.39) was used for $\alpha+{}^3$He.
Using this interaction in MCAS, we
obtained the results listed in Table~\ref{mass7}. 

The parameter values differ (slightly) from those used previously~\cite{Ca06a}
in a study of the same compound  systems but
taken as ${}^3$H and ${}^3$He coupling to an
$\alpha$-particle target. The differences are due primarily to use 
of the nuclear masses listed in ~\cite{Au03} rather than the nucleon mass numbers. 
The comparison of previous with current results is sufficiently good 
to encourage use of the $\alpha$+nucleus program for other systems.

\subsection{The $\alpha+\alpha$ system}

We have also evaluated the spectrum resulting with MCAS for the 
cluster, $\alpha+\alpha$; as another single-channel problem
since the $\alpha$ particles are strongly bound and have no other
bound state in the spectrum.  From~\cite{Fi96}, we note that
the $0_1^+$ and $2_1^+$ states of ${}^8$Be have been found 
with the  $^4$He$(\alpha,\gamma)$ and $^4$He$(\alpha,\alpha)$
reactions.  With a (positive-parity) interaction
[$V_0 = -46.3$ MeV, $V_{ll} = 0.4$ MeV, $R_0 = R_c = 2.1$ fm, and $a_0= a_c = 0.6$ 
fm] with MCAS, two low-excitation resonance states for ${}^8$Be were obtained.
Relative to the cluster threshold, they 
 are the ground state ($0^+$) resonance having centroid and width energies
of 0.095 MeV and 6 eV [c/f experimental values\cite{Ti02} 0.092 MeV
and 5.96 eV] and a first excited ($2^+$) resonance state with 
centroid and width energies of 3.13 MeV and 1.06 MeV
compared with experimental values of 3.03 MeV and 1.51 MeV respectively.
With this simple (local Woods-Saxon) single channel interaction, no 
$4^+$ resonance state is found; at least below 20 MeV excitation.

   No OPP has been used in treating the $\alpha+\alpha$ cluster
 as a single-channel problem since all states found thereby are
 orthogonal.  Thus any state that should be blocked because it requires
 the 8 nucleons to lie in the $0s$-shell simply can be ignored.
 Only if there is channel coupling does a problem arise in ensuring  that
 the Pauli principle is satisfied. With channel coupling, all resultant states of the
 cluster are linear combinations of all states of the same
 spin-parity  defined in the potentials for each of the target states
 considered.  In the present case, the interaction has a $0s$ state bound by
 21 MeV, which, due to Pauli blocking, is deemed to be spurious.

\section{${}^{12}{\rm C}$ as a coupled $\alpha+{}^8{\rm Be}$ system}
\label{carbon}

${}^8$Be is a particle unstable system. The ground and first two
excited states are  $\alpha$-emissive resonances with 
spin-parities and centroid energies of $0^+$ (ground), $2^+$  
at 3.03 MeV, and $4^+$ at 11.35 MeV.  The ground state lies only 
0.092 MeV above the break-up threshold but is a long-enough lived
resonance that in a stellar environment allows capture of a third
$\alpha$ particle  populating  the Hoyle state in ${}^{12}$C
($0^+_2$ state at 7.65 MeV). That state then $\gamma$ decays
to the subthreshold states and, concomitantly, is significantly responsible for
the known abundance of ${}^{12}$C. Thus any acceptable cluster model 
of $\alpha+{}^8$Be $\to {}^{12}$C must find the Hoyle state
correctly located in the spectrum. 

In using MCAS to deduce the low-excitation spectrum of ${}^{12}$C
from an $\alpha+{}^8$Be cluster, we assume that
a collective rotation model prescribes the interactions
of an $\alpha$-particle with each state (of ${}^8$Be) considered,
and forms the coupling between each of those states.
\begin{table}[h]
\begin{ruledtabular}
\caption{\label{Be8-tab}
The potential parameters used for the  interactions in $\alpha+{}^8$Be 
system.  The strengths are in MeV and lengths are in fermi.}
\begin{tabular}{cc|cc}
Potential &  & Geometry/OPP & \\
\hline
$V_0$ & -39.5 & $R_0 = R_c$ & 2.8 \\
$V_{ll}$ &  1.5 & $a_0 = a_c$ & 0.65 \\
$V_{lI}$ & -1.8 & $\beta_2$ & -0.7\\
$V_{II}$ & 2.0 &  $\beta_4$ & 0.2 \\
$V_{mono}$ & -2.7 & $\lambda_s = \lambda_p$ & 10$^6$ \\
\end{tabular}
\end{ruledtabular}
\end{table}
The potential parameters for the collective model Hamiltonian
for the $\alpha+^8$Be cluster are listed in 
Table~\ref{Be8-tab}.  Values used for  
OPP blocking of the $\alpha+{}^8$Be $s$- and $p$-orbits
are listed as well, with the latter only of import in finding
the negative-parity states of the compound.
\begin{figure}[h]
\scalebox{0.7}{\includegraphics*{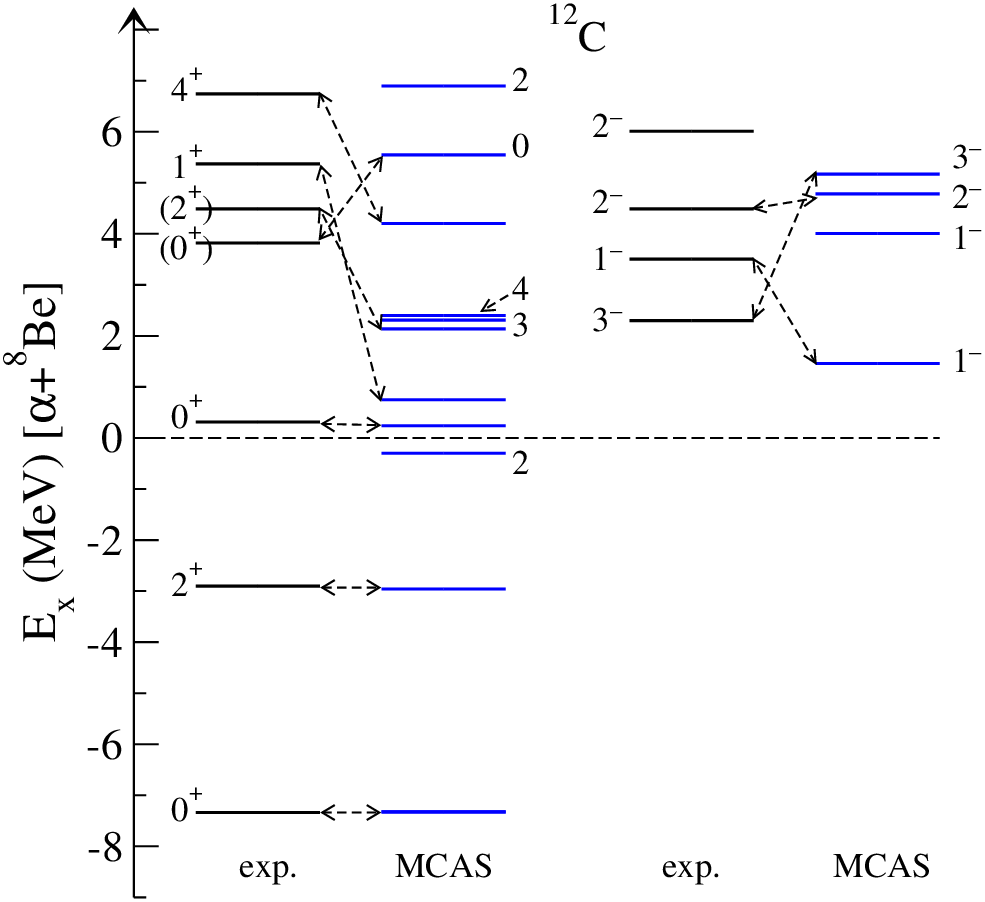}}
\caption{\label{Fig1} (Color online)
The spectrum of low-excitation states in ${}^{12}$C.
The positive-parity results are on the left while
the negative-parity states are shown on the right in this figure.
The MCAS results, found using the interaction 
Hamiltonian listed, is compared to the known set of energies (exp.).}
\end{figure}
The spectrum of positive-parity states in ${}^{12}$C that results is shown in
Fig.~\ref{Fig1}. Energies shown are  relative to the $\alpha+{}^8$Be threshold. 
Clearly there are more evaluated states than in the observed
spectrum  though all those known have calculated equivalents 
in the vicinity of their excitation values.  

With regard to these positive-parity states, the objective
with the simple form for the Hamiltonian was to find the ground,
$2_1^+$, and the Hoyle $0_2^+$ states close to their appropriate energies
relative to the $\alpha+{}^8$Be threshold. The energies of the 
$2_1^+$ and of the $0_2^+$ states set the nuclear interaction  
whose parameters are listed in Table~\ref{Be8-tab}. With that interaction
the ground state was predicted to lie 5.92 MeV below threshold
and to correct that to be 7.37 MeV required addition of a monopole
term in the potential.
Using exactly the same nuclear
interaction determined to match the three low-lying positive-parity 
states of ${}^{12}$C, and with full $p$-orbit blocking,
four negative-parity states were found. Those  
spin-parities were $1^-,2^-$, and $3^-$, as there are in the known spectrum,
but the order, in particular, is not correct. 
By reducing the OPP strengths to have  $p$-orbit
hindrance rather than full blocking, the fully blocked results persist
as long as that strength is greater than $\sim 20$ MeV 
for each target state.

\begin{figure}[h]
\scalebox{0.8}{\includegraphics*{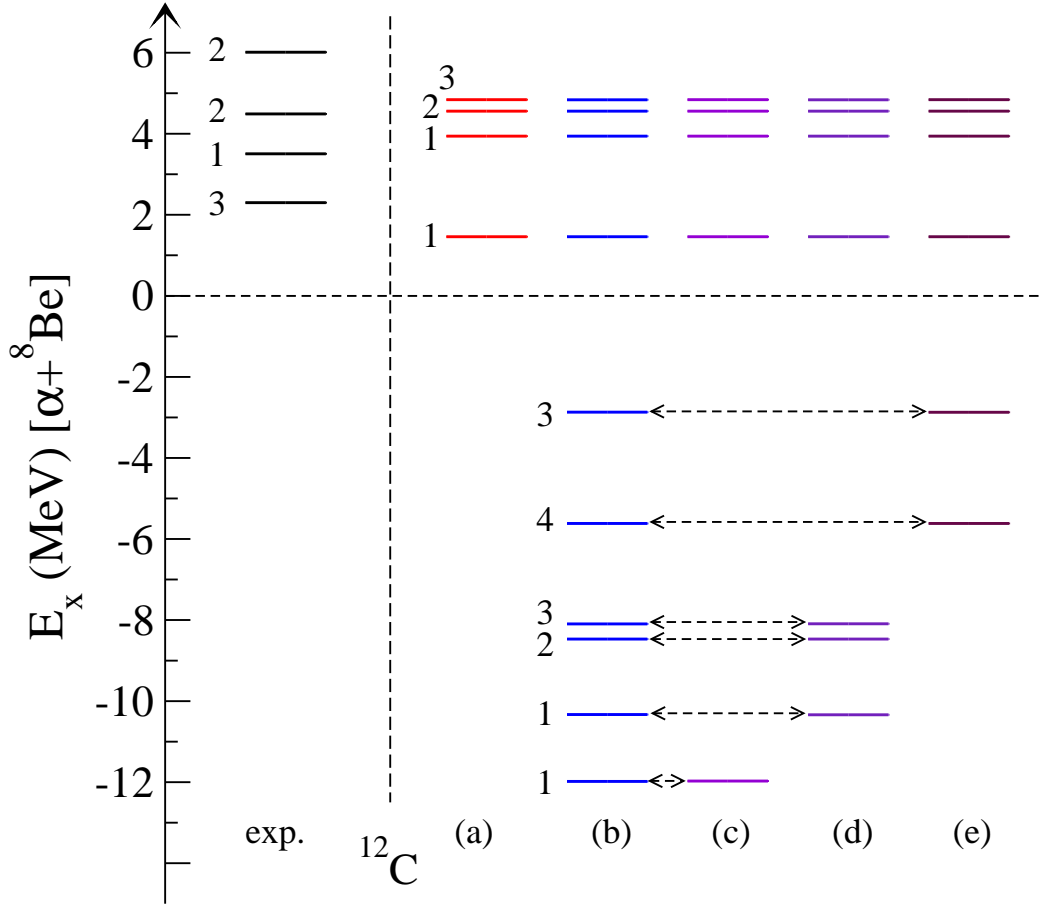}}
\caption{\label{Fig2} (Color online)
The negative-parity spectrum of ${}^{12}$C in relation to the 
$\alpha+{}^8$Be
threshold. The MCAS result found using the interaction Hamiltonian 
listed compared to the known set of state energies (exp.).
Details are given in the text.}
\end{figure}

In Fig.~\ref{Fig2}, the role of the $p$-orbit OPP in 
MCAS results for the low-excitation, isoscalar, negative-parity
states in ${}^{12}$C is displayed.
The energy scale again is taken against the $\alpha+{}^8$Be 
threshold.  The known values
are shown in the column labelled `exp'.
These first MCAS results were found using exactly the same nuclear
interaction determined to match the three low-lying positive-parity 
states of ${}^{12}$C.  The various MCAS results, labelled `(a)' through
`(e)' were found by using diverse OPP blocking of the $p$-orbit in each
of the three target states (of ${}^8$Be) chosen in the calculations.
Full $p$-orbit blocking in all three states gave the results shown in
column (a).  
The MCAS spectrum found when no blocking of the $p$-orbit is made is
shown in column (b), in which there are numerous sub-threshold negative-parity 
states.  The same four resonance states are there though, as they
are in each of the other calculated spectra in this figure.
The results shown in columns (c), (d), and (e) respectively were
obtained by making the $p$-orbit unblocked for the incident $\alpha$ in 
the ground, the $2^+$ state, and the $4^+$ states of the target (${}^8$Be)
respectively.  Without the ground state OPP a deep lying $1^-$ bound
state ensues. Without the $p$-orbit blocking in the $2^+$ state
interactions, a triplet of bound states of spin-parities $1^-, 2^-$ and
$3^-$ is found. While without any such blocking with the target $4^+$
state, extra $3^-$ and $4^-$ bound states appear.

Small variations of the non-central interaction strengths
did not improve the results. In fact,
to get the correct order for the isoscalar, negative-parity resonances in
${}^{12}$C, the negative-parity interaction had to be varied markedly 
from that used to get the positive-parity states. Even then the known
energy gaps were not well reproduced and extra (not observed) states
were found in the low-excitation region considered.

The full set of results are presented in Table~\ref{12C-res} with the 
widths of resonances (in keV) listed along with their energy centroids
(in MeV).
\begin{table}[h]
\begin{ruledtabular}
\caption{\label{12C-res}
Known and MCAS spectral properties of low-excitation states 
in ${}^{12}$C. Widths are in keV, centroids in MeV.}
\begin{tabular}{ccccc}
 $J^{\pi}$ & E$_{exp}$ & $\Gamma_{exp}$ & E$_{MCAS}$ & $\Gamma_{MCAS}$\\
\hline
$0^+$ & -7.37 & --- & -7.37 & --- \\
$2^+$ & -2.93 &  --- & -2.95 & ---\\
$2^+$ & & & -0.30 & ---\\
\hline
$0^+$ & 0.28 & $\sim 0.008$ & 0.24 & 750 \\
$3^+$ & & & 2.29 & 760 \\
$4^+$ & & & 2.43 & 1800 \\
$2^+$ & 3.79 & 430 $\pm$ 80 & 2.15 & 762\\
$1^+$ & 5.34 & 0.02 & 0.75 & 750 \\
$4^+$ & 6.71 & 258 $\pm$ 15 & 4.23 & 770\\ 
\hline
$3^-$ & 2.30 & 35 $\pm$ 5 & 1.81 & 9.2\\
$1^-$ & 3.50& 315 $\pm$ 25 & 2.04 & 1085 \\
$2^-$ & 4.49 & 260 $\pm$ 25 &  &  \\
$2^-$ & 5.98 & 375 $\pm$ 40 & 6.05 & 2800 \\
\end{tabular}
\end{ruledtabular}
\end{table}
The states are listed in the order in which they were found from the
MCAS evaluations and compared with experimental values. 
The widths of the evaluated resonance states are strongly affected 
by the widths of the $2^+$ and $4^+$ states in ${}^8$Be (the experimental
values were used in the calculations).
If one were to treat those as sharp (zero width) then the resultant
resonance widths are many orders of magnitude smaller; some even 
reflecting bound states in the continuum. Strong influence on resonance
widths of a compound nucleus, ${}^9$Be (as $n$+${}^8$Be),
due to the target states themselves being resonances was observed 
using MCAS~\cite{Ca11}.

However, these results from this $\alpha+{}^8$Be study leave much
to be desired and that reflects the difficulty of the model prescription
in which the $\alpha$ coupled to states of the notably deformed 
and unstable target attempting to replicate states of the 
particularly stable compound system, ${}^{12}$C.


\section{${}^{16}{\rm O}$ as a coupled $\alpha+{}^{12}{\rm C}$ system}
\label{oxygen}

Over 35 years ago, semi-microscopic and microscopic $\alpha+{}^{12}$C 
cluster models were used, refs.~\cite{Su76a,Su76b,Ik80,Li80} for example,
to study excited states of ${}^{16}$O. It was found that many  
could be described by $\alpha+{}^{12}$C cluster structures.
Studies of that nature remain  topical as,
in a recent article~\cite{KE14}, cluster structures for positive-parity 
states of ${}^{16}$O were investigated
using the generator coordinate method and an extended $\alpha+{}^{12}$C cluster model. 
The ground and excited states of ${}^{12}$C were taken into account by using wave functions 
defined by the antisymmetrized molecular dynamics method. The $0_2^+, 2_1^+$ and $4_1^+$
states of ${}^{16}$O were described well and those cluster states were found to be
dominated by the $\alpha+{}^{12}$C($0^+_1$) structure.

There are five states in ${}^{16}$O lying below the $\alpha+{}^{12}$C threshold
of 7.16 MeV. All can be populated in ${}^{12}$C($\alpha$,$\gamma$) reactions~\cite{Fi96}.
Thus we anticipated that MCAS evaluations could give estimates of
the  low-lying excitation spectrum of ${}^{16}$O. 

Three states in ${}^{12}$C have been used previously in defining a 
coupled-channel Hamiltonian for the mirror pair, ${}^{13}$C and ${}^{13}$Na,
as nucleon+${}^{12}$C systems. The spectra of those compound 
systems were well reproduced~\cite{Am03}.  The three states used were
the ground $(0^+)$, the first excited $(2^+)$ at 4.44
MeV, and the second excited ($0^+$) at 7.65 MeV.
Low-energy cross sections and analysing powers from the scattering
of nucleons from ${}^{12}$C were also well reproduced~\cite{Am03}.

To study the $\alpha+{}^{12}$C system, we also include
the $3^-$ (9.64 MeV) state in ${}^{12}$C to specify the diverse channels 
in the coupled-channel model used.  Quadrupole
and octupole deformations have been used to specify the coupling
between states and the interaction strengths are listed in Table~\ref{Tab-16O}.
\begin{figure}[h]
\scalebox{0.7}{\includegraphics*{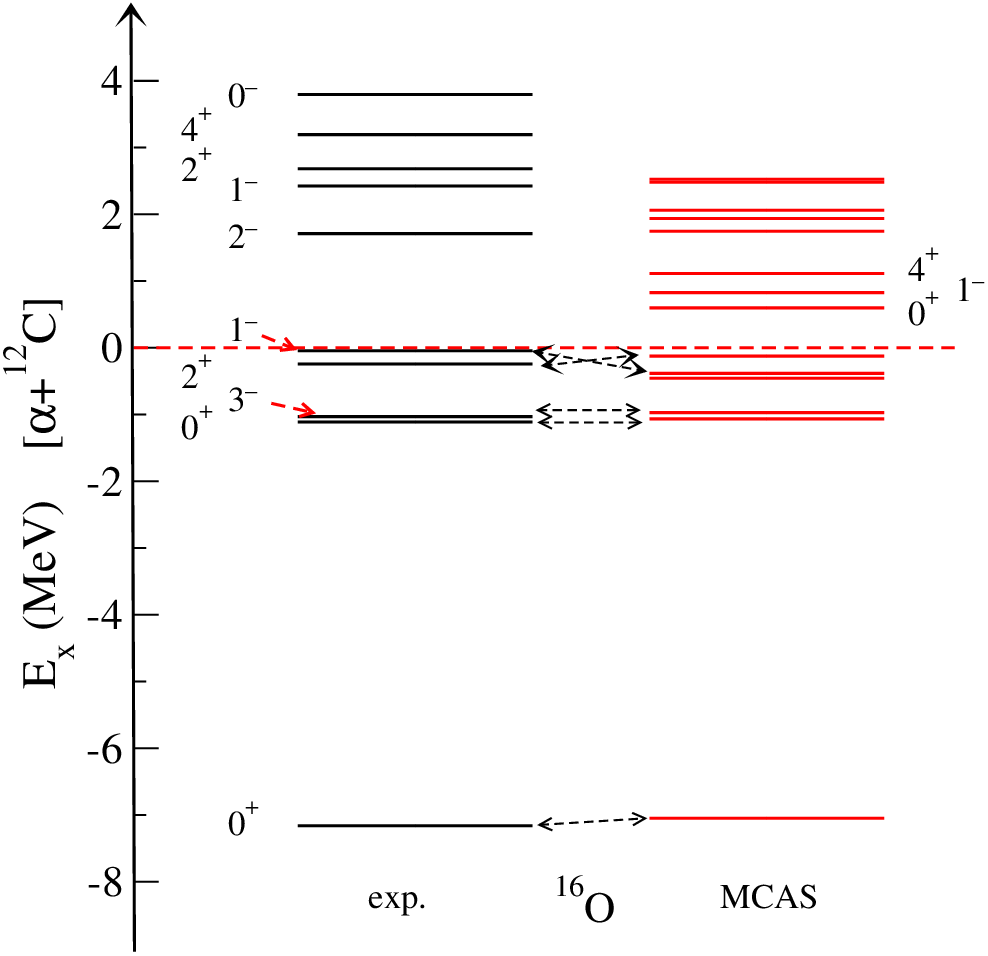}}
\caption{\label{Fig3} (Color online)
The spectrum of ${}^{16}$O in relation to the $\alpha+{}^{12}$C
threshold. The MCAS result found using the interaction Hamiltonian 
listed compared to the known set of state energies (Exp.).}
\end{figure}
\begin{table}[h]
\begin{ruledtabular}
\caption{\label{Tab-16O}
The parameters defining the MCAS Hamiltonian potential for the
$\alpha+{}^{12}$C system. Energies and lengths are in MeV
and fermi respectively.}
\begin{tabular}{lcc}
Type & odd parity & even parity \\
\hline
V$_0$  & $-$27.0 & $-$30.0\\
V$_{\ell\ell}$  & \;\; 1.2 & \;\; 5.0\\
V$_{\ell I}$  & \;\; 1.0 & $-$5.0\\
V$_{II}$  & \;\; 0.0 & \;\; 0.7\\
V$_{mono}$ & --- & $-$3.5\\ 
\hline
Geometry:    & R = R$_c$ = 3.4 & a = a$_c$ = 0.6 \\
\hline
Deformation: & L = 2 & $\beta_n = \beta_c$ = $-$0.52 \\
             & L = 3 & $\beta_n = \beta_c$ = {\phantom{$-$}}0.37\\
\end{tabular}
\end{ruledtabular}
\end{table}
Further, to define a spectrum for
$^{16}$O from the cluster of an $\alpha$ and $^{12}$C
we need to use Pauli blocking and/or Pauli hindrance 
of the  $s$ and $p$-orbits.
The values of the OPP strengths used for that 
are listed in Table~\ref{OPP-16O}. 
\begin{table}[h]
\begin{ruledtabular}
\caption{\label{OPP-16O}
The orthogonalising pseudo-potential strengths in MeV.}
\begin{tabular}{c|cccc}
state & $0^+_{gs}$ & $2^+(4.4389)$ & $0^+(7.654) $ & $3^-(9.641) $ \\
\hline
$s$-wave &  10$^6$ & 10$^6$ & 11.0 & 11.0 \\
$p$-wave &  9.0 & 3.0 & 3.0 & 0.0 \\
\end{tabular}
\end{ruledtabular}
\end{table}

Using the matrix of interaction potentials formed with these parameter
values gives the spectrum for ${}^{16}$O shown in
Fig.~\ref{Fig3}.  The sub-threshold states are in very good
agreement with the known values as they were used to `tune' the 
interaction parameters for both parities. Most notable was the monopole
term since only with that component could the splitting of the ground
to first excited state be determined. Clearly, though, this model fails
to give an adequate spread of resonance states. Nonetheless, each spin-parity
state in the known spectrum (other than the particularly unique 
unnatural-parity $0^-$ state) 
has a matching member in the MCAS result.
\begin{figure}[h]
\scalebox{0.9}{\includegraphics*{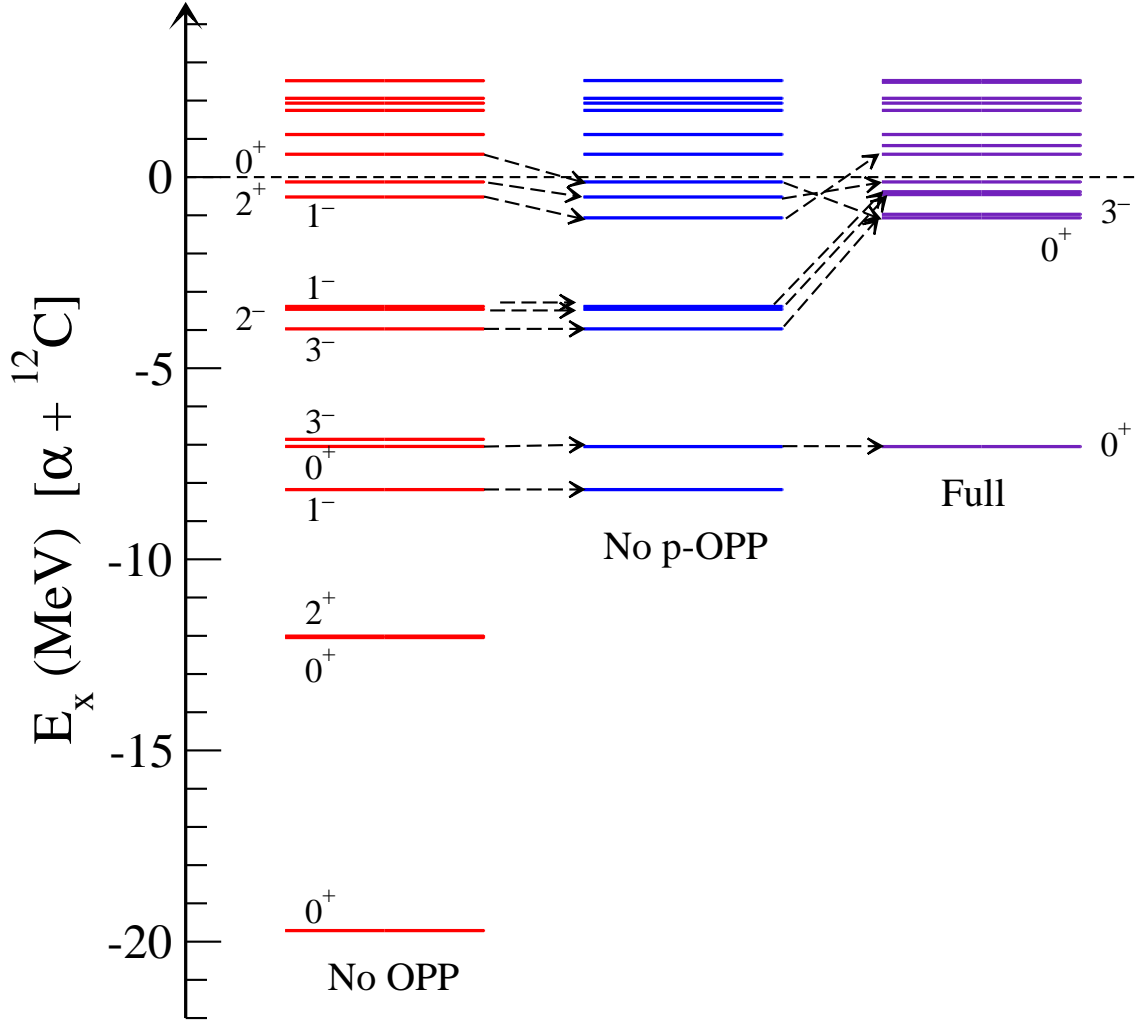}}
\caption{\label{Fig4} (Color online)
MCAS  spectra found using the defined nuclear and Coulomb interactions
and with no OPP included (left), with the $p$-OPP strengths 
in all four states set to zero (middle), compared with the full result 
(right) as given in Fig.~\ref{Fig3}.}
\end{figure}

In Fig.~\ref{Fig4},  the effects of changing the various strength values of the OPP 
entries in the Hamiltonian are shown. With each case, the
specific set and order of the finite number of sturmians used to
expand the interaction matrix of potentials, change. 
Concomitantly the components of the coupled-channel states displayed
in the spectra will also change.
Certainly the admixtures of those components in each compound state
found will vary. Thus the dashed lines connecting the states in this figure
should only be taken as identification of the state spin-parity
in each spectrum.

\section{${}^{20}{\rm Ne}$ as a coupled $\alpha+{}^{16}{\rm O}$ system}
\label{neon}

The $\alpha+{}^{16}$O threshold in ${}^{20}$Ne lies at 4.73 MeV excitation
and just the ground ($0^+$), first excited ($2^+$), and second excited ($4^+$)
states of ${}^{20}$Ne are subthreshold. They are all populated by 
${}^{16}$O($\alpha,\gamma$) processes.   Using MCAS to study the cluster
system seems straightforward but there is the question of how adding an $\alpha$ 
to a system (${}^{16}$O) often thought to have collective attributes
of vibrational model character can achieve a spectrum (of ${}^{20}$Ne) that has   
collective attributes reminiscent of a rotational model.

\begin{figure}[h]
\scalebox{0.7}{\includegraphics*{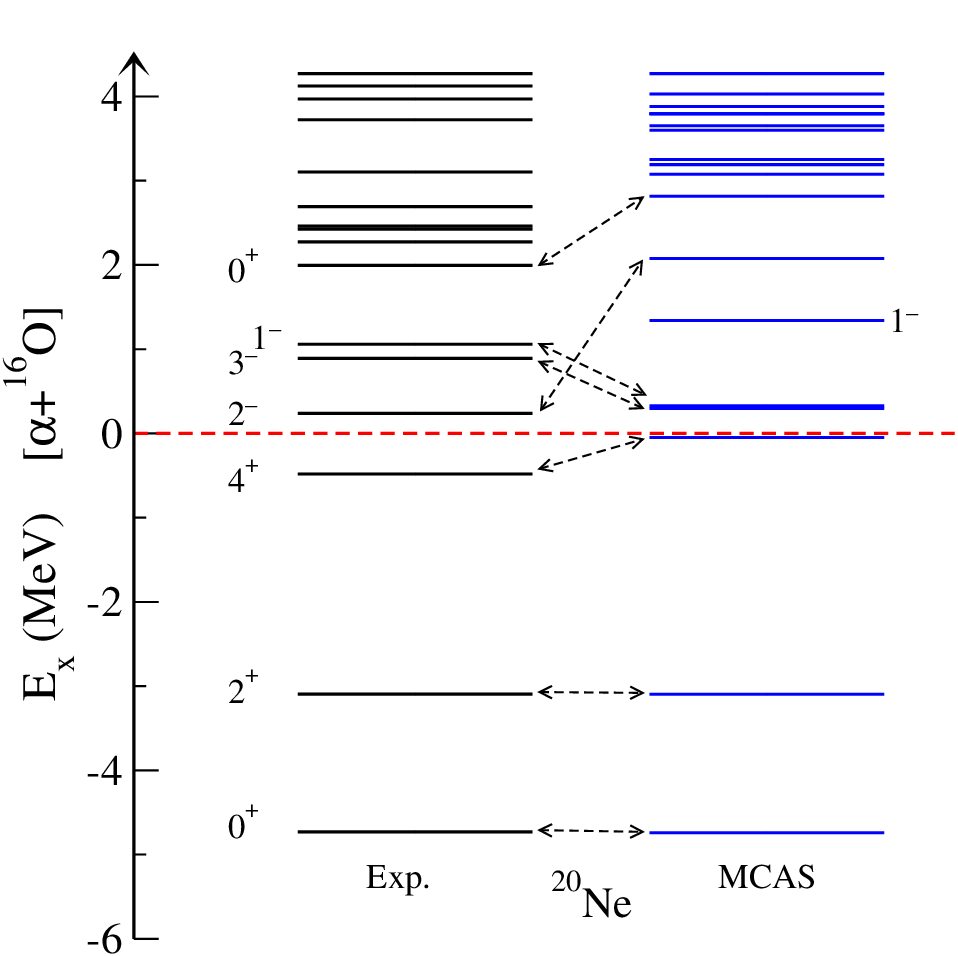}}
\caption{\label{Fig5} (Color online)
MCAS spectrum for ${}^{20}$Ne as an $\alpha+{}^{16}$O cluster
compared to the known one referenced against the $\alpha+{}^{16}$O
threshold.}
\end{figure}
The coupled-channel problem has
been treated with MCAS by considering the interactions with
a vibration model for five low-excitation states in ${}^{16}$O. 
Besides the ground
($0^+$) state treated as the vibration vacuum, we have included
the $0^+_2$ (6.05 MeV) state as a two quadrupole phonon excitation, 
the $3^-$ (6.13 MeV) state as a single octupole excitation,
the $2^+$ (6.92 MeV) state as a single quadrupole excitation,
and the $1^-$ (7.12 MeV) state as a two phonon (quadrupole plus
octupole) excitation. We also assumed that the bound $s$ and $p$-orbits 
were blocked by using the OPP weights,
$\lambda_s = \lambda_p = 10^6$ MeV.
\begin{table}[h]
\begin{ruledtabular}
\caption{\label{Tab-20Ne}
The parameters defining the MCAS Hamiltonian potential for the
$\alpha+{}^{16}$O system. Energies and lengths are in MeV
and fermi respectively.}
\begin{tabular}{lccc}
Type & Negative & Positive & Geometry \\
 V$_0$ & $-$26.0 & $-$27.2 & R = R$_c$ = 3.4 \\ 
V$_{\ell\ell}$ & \ 0.16 & \ 0.16 & a = a$_c$ = 0.65 \\
V$_{II}$  & $-$0.19 & $-$0.19 & L = 2\ ;\  $\beta_n = \beta_c$ = 0.52 \\
V$_{mono}$ & & $-$0.23 & L = 3\ ;\  \hspace*{1.4cm} = 0.4 \\
\end{tabular}
\end{ruledtabular}
\end{table}
Then on using the parameter set listed in Table~\ref{Tab-20Ne},
the resulting  spectrum is shown in Fig.~\ref{Fig5} in comparison with the known
one.  In this case, no $V_{\ell I}$ component was needed to get
the spectrum labelled MCAS in the figure.

The seven lowest excitation states in the known spectrum have
matching entries in the MCAS result and,
with the exception of the unnatural-parity $2^-_1$ state,
they agree to better than an MeV. The evaluated $4^+_1$ state  is just
bound while the low-lying $3^-$ and $1^-$ resonant states are in the observed
order but their centroid energies are close and smaller than is observed.
Also the calculated spectrum has an additional $1^-$ resonance whose  centroid
is at 1.34 MeV. The known $1^-_2$ resonance centroid is at 4.12 MeV
in this figure.

The full set of results are presented in Table~\ref{20Ne-res} with the 
widths of resonances (in keV) listed along with their energy centroids
(in MeV).
\begin{table}[h]
\begin{ruledtabular}
\caption{\label{20Ne-res}
Known and MCAS spectral properties of low-excitation states 
in ${}^{20}$Ne. Widths are in keV, centroids in MeV.}
\begin{tabular}{ccccc}
 $J^{\pi}$ & E$_{exp}$ & $\Gamma_{exp}$ & E$_{MCAS}$ & $\Gamma_{MCAS}$\\
\hline
$0^+_1$ & -4.73 & --- & -4.74 & --- \\
$2^+_1$ & -3.09 & --- & -3.10 & ---\\
$4^+_1$ & -0.48 & --- & -0.05 & ---\\
\hline
$2^-_1$ & 0.24 & $\gamma$-decay & 2.08 & $< 10^{-13}$\\
$3^-_1$ & 0.89 &  $\pm$  & 0.30 & $3 \times 10^{-13}$\\
$1^-_1$ & 1.09 &  (2.8 $\pm$ 0.3) 10$^{-2}$ & 0.33 & $<10^{-13}$ \\
$0^+_2$ & 2.00 & 19 $\pm$ 0.9 & 2.82 & 230 \\
$1^-_2$ & 3.98 & 19 & 1.34 & 0.32 \\
\hline
$4^-_1$ & 2.27 & $\gamma$-decay & 3.88 & $< 10^{-13}$\\
$3^-_2$ & 2.43 & 8.2 $\pm$ 0.3 & 3.19 & 4 \\
$0^+_3$ & 2.46 & 3.4 $\pm$ 0.2 & 3.08 & 820 \\
$2^+_2$ & 2.69 & 15.1 $\pm$ 0.7 & 3.65 & 228 \\
$2^+_3$ & 3.10 & 2 & 4.03 & 105 \\
$5^-_1$ & 3.72 & 0.013 & & \\
$0^+_4$ & $\sim 4$ & $> 800$ & 3.25 & 40 \\
\end{tabular}
\end{ruledtabular}
\end{table}
With one exception, MCAS finds all states additional to those labelled explicitly in
Fig.~\ref{Fig5}  to within an MeV of the tabulated data.
The exception is the $4^-_1$ state but as this only decays
by $\gamma$ emission it may not be well described as an $\alpha$
cluster model.  The widths of the resonances are not well matched by the
simple model calculation and that points to the need for a better
prescription of the coupled-channel Hamiltonian.


\section{MCAS and elastic scattering of $\alpha$ particles}
\label{elscat}

Besides producing bound and resonant state expectations for the
spectrum of the compound system studied, MCAS can determine
scattering matrices for energies above the particular cluster
threshold, and thus, cross sections. In the past, for the compound systems
${}^{13}$C (treated as $n$+${}^{12}$C)~\cite{Am03}, and ${}^{15}$F 
(treated as $p$+${}^{14}$O)~\cite{Ca06}, scattering cross sections were obtained 
that were in very good agreement with measured data. 
Even spin-dependent measurable data were matched in the former, and 
resonances were predicted with the latter, some of which 
were subsequently discovered~\cite{Mu09,Mu10}.

Much data has been taken for low-energy scattering of $\alpha$ particles, from ${}^{12}$C
and ${}^{16}$O in particular, and resonance features are very evident in
the associated elastic scattering cross sections. 
That is especially the case  at backward scattering angles.
We show just one example herein, namely that of elastic scattering of 
$\alpha$ particles from ${}^{16}$O taken at 165$^\circ$ and for energies
between 2 and 6 MeV, simply to illustrate the utility of MCAS to 
determine scattering cross sections.

\begin{figure}[h]
\scalebox{0.7}{\includegraphics*{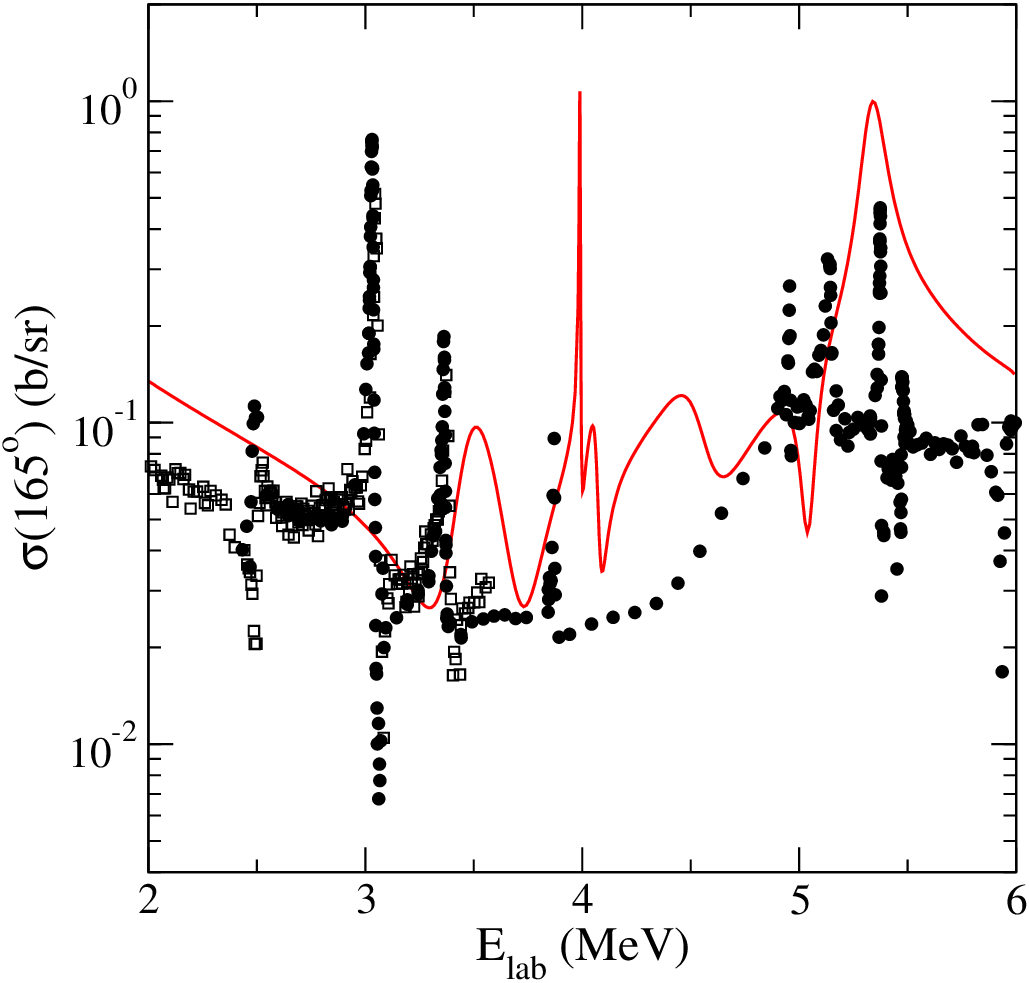}}
\caption{\label{Fig6} (Color online)
Cross sections for $\alpha+{}^{16}$O elastic scattering at 165$^\circ$ 
for a range of incident $\alpha$-particle energies.
The MCAS result (solid curve) is compared with two data sets;
the filled circles are the data of ~\cite{De06} while the open squares are those
of~\cite{Ja85}.}
\end{figure}

Two data sets and our MCAS result are shown in Fig.~\ref{Fig6}.
Clearly the known detailed structure is not reproduced, but that was not unexpected 
since, as yet, the Hamiltonian does not give resonant states of the 
compound (${}^{20}$Ne) 
sufficiently in agreement with the known spectrum.   
That is the case with $\alpha+{}^{12}$C scattering as well, 
and for the same reason.
However, with the example shown, the average magnitude of the 
cross section and of the evaluated resonant structures
in the energy range are comparable  in magnitude and  width with those in the data.  
These results serve encourage to find improved model coupled-channel 
interactions, to those from the quite simple collective models we have used to date,
to find better evaluated  spectra of the compound systems.

\section{Conclusions}
\label{conclusion}

The MCAS method has been used to define
low-energy spectra for  ${}^{12}$C, ${}^{16}$O, and ${}^{20}$Ne
with the systems considered as the coupling of an $\alpha$ particle with
low-excitation states of the core nuclei, ${}^8$Be, ${}^{12}$C, and
${}^{16}$O, respectively.  Collective models have been used to define
the matrices of interacting potentials and quadrupole (and octupole when
relevant) deformation allowed and taken to second order.
The program was checked by it replicating results found previously
with a version built to describe spin-$\frac{1}{2}$ nuclei clustered
with nuclei having ground state spin-parity $0^+$. Results of the
single-channel $\alpha+\alpha$ cluster found the essential known states 
of ${}^8$Be in good agreement with the known spectrum.

Treating ${}^{12}$C as an $\alpha+{}^8$Be cluster required a 
monopole term (strength -2.7 MeV) to achieve a good replication of the 
ground, $2^+$ (4.44 MeV), and the just unbound $0^+_2$ (7.65 MeV) states;
the extremely narrow width of that $0^+_2$ was not matched however.
All other known resonance states to $\sim 12$ MeV were found to have partners
in the MCAS spectrum but without a close match. 
The results reflected the difficulty of using a simple model cluster
prescription in seeking a spectrum of a particularly stable compound
nucleus when one of the elements is a strongly deformed and unbound nucleus.
Treating ${}^{16}$O as a 
cluster of an $\alpha$ with the quite stable ${}^{12}$C led to a good
match for the sub-threshold (for $\alpha$ breakup) bound states of the
compound.  Again a monopole term (strength -3.5 MeV) was required to 
depress the ${}^{16}$O ground state below the next set of its excited ones.
Resonance states found with MCAS match spin-parities of the quintuplet of 
resonances in the known spectrum but with energy centroids $1 - 2$ MeV
too small.

We considered ${}^{20}$Ne as a coupled-channel problem of an $\alpha$
with the low-excitation spectrum of ${}^{16}$O. The interactions in this case
were specified using a pure vibration model for the core nucleus. In this study
the role of Pauli blocking of some $\alpha$-particle states from the 
coupled-channel problem was essential in finding the resultant spectrum
in agreement with the known one, having three subthreshold states with
energies reminiscent of a rotation model. The lowest known resonance states
have partners in the MCAS result though, once more, the energy centroids
are $1 - 2$ MeV from matching and the widths found are not in good 
agreement with the tabulated ones. 

Finally, as an example of the utility in using MCAS to define scattering cross
sections, an MCAS result for low-energy $\alpha$-particle elastic scattering from
${}^{16}$O was compared with data taken at 165$^\circ$.  The same multi-channel 
interaction used to get the spectrum of ${}^{20}$Ne was used and so observed 
resonance states were not matched.  Nonetheless the average
magnitude of the evaluated cross section is comparable to the measured one
and  resonance shapes and magnitudes obtained using MCAS are similar to the measured
ones save that they are not at the observed energies. These characteristics 
we can expect to improve as better model interactions are used in the
coupled-channel Hamiltonian.

Each of the systems studied has some aspects that concur with the known compound
system low-excitation spectrum when a monopole term is included in 
the interaction matrix of potentials.  This extra term we believe is a
reflection of strong pairing in the ground states of the $N=Z$ compound
nuclei considered. Each cluster has unique problematic aspects for
treatment with the simple collective model prescriptions for the 
interactions of an $\alpha$ particle with the core nuclei. 
$^8$Be is an unbound system not only
having resonance states but it is also strongly deformed.  ${}^{12}$C 
has a rotor like low-excitation spectrum but the compound, ${}^{16}$O,
has a low-excitation spectrum reminiscent of a vibrations on a spherical 
ground state.  Then 
the $\alpha+{}^{16}$O cluster seeks to form ${}^{20}$Ne whose low-excitation
spectrum seems to have rotor like characteristics. 
Nonetheless, with reasonable interactions and orthogonalising
pseudo-potentials accounting for credible Pauli blocking of the the relative
motion of nucleons in the $\alpha$+nucleus  clusters, these most simple prescriptions
of the coupled-channel problems give good descriptions of sub-threshold states
in the compound systems, and of some resonance states. This makes further
investigations with MCAS leading to assessments of low-energy $\alpha$-capture
processes worthwhile.

\begin{acknowledgments}
JPS acknowledges support from the Natural Sciences and Engineering Research Council of
Canada.  SK acknowledges support from the National Research Foundation of South Africa. 
PRF acknowledges funds from the Dipartimento di Fisica e Astronomia
dell'Universit di Padova and the PRIN research project 2009TWL3MX,
and of helpful conversations with Peter O. Hess.
\end{acknowledgments}

\appendix

\section{Collective models of $\alpha$+nucleus potential matrices}
\label{app1}

For an $\alpha$+nucleus cluster system, the symmetrized form of the matrix 
of interaction potentials to be used is 
\begin{align*}
V_{cc'}(r) &=\ V_0 f_{cc'}(r) 
+ \frac{1}{2} V_{\ell \ell} \sum_{c''} 
\bigg( [\ell \cdot \ell]_{cc''} f_{c''c'}(r)  
+ f_{cc''}(r) [\ell \cdot \ell]_{c''c'} 
\bigg)\
\\
&\hspace*{2.5cm}
+ \frac{1}{2} V_{II} \sum_{c''} 
\bigg( [{\bf I \cdot I}]_{cc''} f_{c''c'}(r)  
+ f_{cc''}(r) [{\bf I \cdot I}]_{c''c'} 
\bigg)\
\\
&\hspace*{2.5cm}
+ \frac{1}{2} V_{\ell I} \sum_{c''} 
\bigg( [\ell {\bf \cdot I}]_{cc''} g_{c''c'}(r)  
+ g_{cc''}(r) [\ell \cdot {\bf I}]_{c''c'} 
\bigg)\
\\
&=\  
\left[V_0 
+ \frac{1}{2} V_{\ell \ell} \bigg( 
\ell' (\ell' + 1) + \ell (\ell + 1)
\bigg)
+ \frac{1}{2} V_{II} \bigg( 
I' (I' + 1) + I (I + 1)
\bigg) \right] 
\ f_{cc'}(r)
\\
&\hspace*{0.6cm} 
+ \frac{1}{4} V_{\ell I} \bigg( 2 J(J + 1)  
- \ell' (\ell' + 1) - \ell (\ell + 1)
- I' (I' + 1) - I (I + 1)
\bigg) \ g_{cc'}(r) ,
\label{Veqn}
\end{align*}
since,  
\begin{align}
[\ell \cdot \ell]_{cc'} 
&=\ \sum_{m_\ell m_{\ell'} N N'} 
\left< \ell I m_\ell N \vert J M \right>
\left< \ell' I' m_{\ell'} N' \vert J M \right>
\ <\ell m_\ell \left| {\vec{l}^2} \right| \ell' m_{\ell'} > 
\delta_{I'I} \delta_{N'N}
\nonumber\\
&=\ \ell (\ell + 1) \sum_{m_\ell N}
\left[ 
\left< \ell I m_\ell N \vert J M \right>
\right]^2 \delta_{I'I} \delta_{N'N}
\delta_{\ell' \ell} \delta_{m_\ell m_{\ell'}}
=\ \ell (\ell + 1) \delta_{cc'} ,
\end{align}
and likewise $[{\bf I\cdot I}]_{cc'} = I(I + 1)\delta_{cc'}$.
Similarly one finds that, for $\alpha$+nucleus systems as 
$\ell + {\bf I} = {\bf J}$ is conserved,
\begin{equation}
[\ell \cdot {\bf I}]_{cc''} \ 
=\ \frac{1}{2} \left[ J (J + 1) - \ell (\ell + 1) - I (I + 1)
\right] \delta_{c c''}\ .
\end{equation}

\section{Using a rotation model of the target}
\label{app2}

The surface of a rigid drop of matter, with permanent, but axially symmetric,
deformation from the spherical, is represented by the expansion
\begin{equation}
R = R_0 \left[ 1 + \sum_L c_L P_L(\theta') \right] ,
\end{equation}
where the angles refer to a body fixed symmetry axis, $\phi'$ being understood.  
The $C_L$ are suitable coefficients to form the chosen shape of the 
surface of the nucleus.
This transforms to a space-fixed frame to take the form
\begin{equation}
R = R_0 \left[ 1 + \sum_{L} \sqrt{\frac{4\pi}{(2L + 1)}} \
\beta_L \ \left[{\bf Y}_L(\Omega) {\bf \cdot Y}_L(\xi)\right] \right]\ ,
\label{Rrotor}
\end{equation}
where $\Omega (\theta \phi)$ refer to the space-fixed axis and $\xi$ are the 
Euler angles of the transformation and $\beta_L$ are the usual deformation
parameters.  Even with odd-mass nuclei for which the states will have 
half-integer spin, we will presuppose that collective excitation is of the 
underlying even-mass core.

A rotor model prescription for $f(r)$, with deformation
\begin{equation}
\epsilon = \sum_{L} \sqrt{\frac{4\pi}{(2L + 1)}} \
\beta_L \ \left[{\bf Y}_L(\Omega) {\bf \cdot Y}_L(\xi)\right] 
\end{equation}
taken to second order, gives a channel-space matrix
\begin{align}
f_{cc'}(r) =&\ \bigg\{ f_0(r) 
- R_0 
\ \frac{df_0(r)}{d\epsilon}
\sum_{L} \sqrt{\frac{4\pi}{(2L + 1)}} \
\beta_L \ \left[{\bf Y}_L(\Omega) {\bf \cdot Y}_L(\xi)\right]
\nonumber\\
&\hspace*{1.0cm}
+ R_0^2 \frac{d^2f_0(r)}{d\epsilon^2}
\sum_{L_1, L_2} \frac{4\pi}{\sqrt{(2L_1 + 1)(2L_2 + 1)}} \
\beta_{L_1} \beta_{L_2} 
\nonumber\\
&\hspace*{5.5cm}\times
\left[{\bf Y}_{L_1}(\Omega) {\bf \cdot Y}_{L_1}(\xi)\right]
\ \left[{\bf Y}_{L_2}(\Omega) {\bf \cdot Y}_{L_2}(\xi)\right]
\bigg\}_{cc'} 
\nonumber\\
=&\ [f_0(r)]_{cc'} 
- R_0\ \frac{df_0(r)}{dr}
\sum_L \sqrt{\frac{4\pi}{(2L+1)}}\ \beta_L 
\left[{\bf Y}_L(\Omega_r) {\bf \cdot Y}_L(\xi) \right]_{cc'}
\nonumber\\
&\hspace*{1.0cm}  + \frac{1}{2} R_0^2 \frac{d^2f_0(r)}{dr^2}
\left[ \sum_{L_1 L_2}\ \sqrt{(2L_1+1)(2L_2+1)} 
\ \beta_{L_1}\ \beta_{L_2} \right.
\nonumber\\
&\hspace*{2.5cm}\times \left.
\sum_{\cal L}\frac{1}{(2{\cal L}+1)}
\left[ 
\left< L_1 L_2 0 0 \vert {\cal L} 0\right> 
\right]^2\ 
\left[{\bf Y}_{\cal L}(\Omega_r) {\bf \cdot Y}_{\cal L}(\xi) \right]_{cc'}
\right]  ,
\label{feqn}
\end{align}
where we have used the identity
\begin{align}
&\left[{\bf Y}_{L_1}(\Omega) {\bf \cdot Y}_{L_1}(\xi) \right]\
\left[{\bf Y}_{L_2}(\Omega) {\bf \cdot Y}_{L_2}(\xi) \right]
\nonumber\\
&\hspace*{0.7cm} =
\frac{1}{4\pi}\  (2L_1 + 1)\ (2L_2+1)
\sum_{{\cal L}}
\frac{1}{2{\cal L} + 1}
\left[
\left< L_1 L_2 0 0 \vert {\cal L} 0 \right>
\right]^2
\left[{\bf Y}_{{\cal L}}(\Omega) {\bf \cdot Y}_{{\cal L}}(\xi) \right].
\label{tenprod}
\end{align}

Then the potentials are 
\begin{align*}
V_{cc'}(r) &=\  
\left[ \bigg( V_0 + V_{\ell \ell} \ell (\ell + 1)
+ V_{II} I(I+1) \bigg) f_0(r) \right.
\\
&\hspace*{3.0cm}\left.
+ \frac{1}{2} V_{\ell I}(r) \bigg( J (J + 1) - \ell (\ell + 1) - I (I+1)
\bigg) g_0(r) \right] \delta_{cc'}
\\
&\hspace*{0.5cm}
 - R_0 \Biggl[ \frac{df_0(r)}{dr}
\bigg( V_0 + 
\frac{1}{2} V_{\ell \ell} \left[
\ell' (\ell' + 1) + \ell (\ell + 1) \right] 
+ \frac{1}{2} V_{II} \left[
I' (I' + 1) + I (I + 1) \right] 
\bigg) 
\\
&\hspace*{0.9cm} 
+ \frac{1}{4} \frac{dg_0(r)}{dr} V_{\ell I}
\bigg( 2J(J+1) - \ell' (\ell' + 1) - \ell (\ell + 1) -I'(I'+1) - I (I+1)
\bigg) \Biggr]
\\
&\hspace*{2.5cm}\times 
\sqrt{(2\ell' + 1)}\
\sum_L\ \beta_L \ 
(-1)^{(\ell' + I +J)}\
\left< \ell' L 0 0 \vert \ell 0\right>
\left\{
\begin{array}{ccc}
I & L & I'\\
\ell' & J & \ell
\end{array}
\right\}
\\
&\hspace*{9.0cm}\times
< I' \left\| {\bf Y}_L \right\| I >\\
& + \frac{1}{2} R_0^2 
\Biggl[
\frac{d^2f_0(r)}{dr^2}
\bigg(V_0 
+ \frac{1}{2} V_{\ell \ell} \left[
\ell' (\ell' + 1) + \ell (\ell + 1) 
\right] 
+ \frac{1}{2} V_{II} \left[
I' (I' + 1) + I (I + 1) 
\right] 
\bigg)
\nonumber\\
&\hspace*{0.9cm} 
+ \frac{1}{2} \frac{d^2g_0(r)}{dr^2} V_{\ell I} 
\bigg( 2J(J+1) - \ell' (\ell' + 1) - \ell (\ell + 1) -I'(I'+1) - I (I+1)
\bigg) \Biggr]
\nonumber\\
&\hspace{2.0cm}\times 
\ (-1)^{(\ell' + I + J)}\ \frac{1}{\sqrt{4\pi}}
\sqrt{(2\ell' + 1)}\ 
\sum_{L_1 L_2} \beta_{L_1}\ \beta_{L_2}\ \sqrt{(2L_1+1)(2L_2+1)} 
\nonumber\\
&\hspace*{2.5cm}\times
\sum_{\cal L}\ \frac{1}{\sqrt{(2 {\cal L} + 1)}}
\left[ 
\left< L_1 L_2 0 0 \vert {\cal L} 0\right>
\right]^2\ 
\left< \ell' {\cal L} 0 0 \vert \ell 0\right>
\left\{
\begin{array}{ccc}
I & {\cal L} &  I'\\
\ell' & J & \ell
\end{array}
\right\}
\nonumber\\
&\hspace*{9.0cm}\times
< I \left\| {\bf Y}_{\cal L} \right\|I' > .
\label{result1}
\end{align*}
Note we have retained all terms in the phase factor since $I'$ may be
integer of half-integer.  

To find the structure factors in the above, consider the
Hamiltonian for a general quantum rotor to be 
\begin{equation}
H_{total} = H + H_{intrinsic}
\hspace*{0.5cm};\hspace*{0.5cm}
H = H_{rot} = \frac{\hbar^2}{2}\left[
\frac{1}{{\cal I}_1} L_1^2
\ +\ \frac{1}{{\cal I}_2} L_2^2
\ +\ \frac{1}{{\cal I}_3} L_3^2
\right] ,
\end{equation}
where the moment of inertia are about body fixed axes with 3
taken to be the equivalent to the space-fixed $z$-axis.
An intrinsic Hamiltonian has been included though we will
assume that the intrinsic state does not change with the 
low-excitation states to be considered. 

The basic eigenvectors $|LMK\rangle$ satisfy
\begin{equation}
{\bf L}^2 |LMK\rangle = L (L+1) |LMK\rangle
\ ,\hspace*{0.3cm}
L_z |LMK\rangle = M |LMK\rangle
\ ,\ {\rm and}\hspace*{0.3cm} 
L_3 |LMK\rangle = K |LMK\rangle .
\end{equation}
The adiabatic condition is assumed so that the intrinsic 
and rotational degrees of freedom decouple.
Then, for a rotor in general having  no axis of symmetry,
its eigenstates will have the form
\begin{equation}
|IM\rangle = \sum_{K=-I}^I A_k |IMK\rangle .
\end{equation}
However, there are symmetry conditions that cause the 
general Hamiltonian to have restrictions giving two groups of
general solutions: those that 
have $K$ quantum numbers all even and those that have them all 
odd. Additionally there are invariances as to the specific labelling of
the body fixed axes.
We consider the excited states for use in MCAS evaluations as
being members of the collective model sets with lowest 
energies. This usually restricts consideration to the 
A-representation~\cite{Da68,Ro70} (for positive-parity states),
whence $A_{-K} = (-)^I A_K$.  If negative-parity states are 
required, then members of the B$_1$-representation,
for which  $A_{-K} = (-)^{I+1} A_K$,
need be considered.
For most even-mass nuclei, the ground state spin-parity is $0^+$
and we restrict consideration to nuclear systems having  axial
symmetry.  

The nuclear states considered then are eigenfunctions of the quantised 
rotor Hamiltonian
\begin{equation}
H = \frac{\hbar^2}{2}\left[
\frac{1}{{\cal I}_0} L_1^2
\ +\ \frac{1}{{\cal I}_0} L_2^2
\ +\ \frac{1}{{\cal I}_3} L_3^2
\right]\
=\ \frac{\hbar^2}{2}\left[
\frac{1}{{\cal I}_0} {\mathbf L}^2
+
\left( \frac{1}{{\cal I}_3}
 - \frac{1}{{\cal I}_0} \right) \ L_3^2
\right] ,
\end{equation}
which has eigenenergies given by
\begin{equation}
H |LMK\rangle = \frac{\hbar^2}{2} \Bigg[
\frac{1}{{\cal I}_0} L(L+1)\ +\
\bigg(\frac{1}{{\cal I}_3} - \frac{1}{{\cal I}_0} \bigg)
K^2 \Bigg]
|LMK> .
\end{equation}
As we limit consideration to axial symmetric cases, we must impose
invariance under rotation of 180$^\circ$ about any axis perpendicular to
the symmetry axis.  With $|IMK\rangle = D^I_{MK}(\xi)$ and incorporating the 
eigenstates of the intrinsic Hamiltonian~\cite{Da68,Ro70},
\begin{equation}
H_{intrinsic}\ \phi_K(\omega) = \epsilon_0(K) \ \phi_K(\omega),
\end{equation}
the normalised eigenstates of interest are
\begin{equation}
\Psi_{I, MK}(\xi,\omega) = 
\sqrt{\frac{(2I+1)}{16 \pi^2 \left(1 + \delta_{K 0}\right)}}
\ \Bigg[ D^I_{MK}(\xi) \phi_K(\omega) + (-)^{(I+K)} D^I_{M, -K}(\xi)
\phi_{\bar K}(\omega) \Bigg].
\label{PsiMK}
\end{equation}
Here $\bar K$ is defined from the symmetry requirement 
\begin{equation}
\phi_{\bar K}(\omega) = {\cal R}_2(-\pi)\ \phi_K(\omega)
= e^{i\pi J_2} \phi_{K}(\omega) = \pm \phi_{K}(\omega)
\hspace*{1.0cm} {\rm as\ I\ is\ even/odd}. 
\label{roteq}
\end{equation}
Let $r_I (=\pm 1)$ be the eigenvalues of Eq.~(\ref{roteq}). 

In the specification of the coupled-channel potentials for MCAS, 
reduced matrix elements of $Y_{LM}^\star$,
with angles relating to the body-fixed axes, are required. 
Furthermore we assume that the intrinsic state does not change
for low-excitation spectra.
Matrix elements with states given in Eq.~(\ref{PsiMK}) then are
\begin{align}
&< \Psi^I_{NK} \left| Y_{LM}^\star \right|
\Psi^{I'}_{N'K'} >
\ =\ \frac{1}{16\pi^2} 
\sqrt{\frac{(2I+1)(2I'+1)}{(1+\delta_{K 0})(1+\delta_{K' 0})}} \ 
\Bigg\{
< D^I_{NK} \left| Y_{LM}^\star \right| D^{I'}_{N'K'} >
\ \delta_{K K'}
\nonumber\\
&\hspace*{3.5cm}
+ (-)^{(I+I'+K+K')}
< D^I_{N,-K} \left| Y_{LM}^\star \right| D^{I'}_{N',-K'} >
\ r_I\ r_{I'}\ \delta_{{\bar K} {\bar K'}}
\nonumber\\
&\hspace*{3.5cm}
+ (-)^{(I+K)}
< D^I_{N, -K} \left| Y_{LM}^\star \right| D^{I'}_{N'K'} >
\ r_I\ \delta_{{\bar K} K'}
\nonumber\\
&\hspace*{3.5cm}
+ (-)^{(I'+K')}
< D^I_{NK} \left| Y_{LM}^\star \right| D^{I'}_{N', -K'} >
\  r_{I'}\ \delta_{K{\bar K}}
\Bigg\} .
\end{align}
As
\begin{align}
&Y_{LM}^\star(\beta, \alpha)\ =\ \sqrt{\frac{(2L+1)}{4\pi}}\
 D^L_{M0}(\alpha, \beta, \gamma)
\nonumber\\
&< D^I_{NK} \left| D^L_{M0} \right| D^{I'}_{N'K'} >
\ =\ 
\frac{8 \pi^2}{(2I+1)}\ 
\left< I' L N' M | I N\right>\  
\left< I' L K' 0 | I K\right>\  \delta_{K' K},
\label{meDs}
\end{align}
the matrix elements reduce to
\begin{align}
&< \Psi^I_{NK} \left| Y_{LM}^\star \right|
\Psi^{I'}_{N'K'} >
\ =\ \sqrt{\frac{(2I'+1) (2L+1)}{4\pi (2I+1)}} 
\left< I' L N' M | I N\right>\  
\nonumber\\
&\hspace*{0.5cm}
\times \frac{1}{2}\ \frac{1}{(1 + \delta_{K0})} 
\Bigg\{
\left< I' L K 0 | I K\right> 
+ (-)^{(I + I')} 
\left< I' L -K 0 | I -K\right>\ r_I r_{I'} 
\nonumber\\
&\hspace{1.5cm}
+ (-)^{(I + K)} \left< I' L -K 0 | I K\right>\ r_{I'} 
+ (-)^{(I' + K)} \left< I' L K 0 | I -K\right>\ r_I 
\Bigg\} .
\label{eq5-13}
\end{align}
The latter two terms contribute only for $K = 0$ and so
offset the factor $(1 + \delta_{K0})^{-1}$.
The reduced matrix elements are then identified by
\begin{align}
&< \Psi^I_{K} \left\| Y_{L} \right\| \Psi^{I'}_{N'} >
=  
\frac{1}{2}\ \sqrt{\frac{(2I'+1) (2L+1)}{4\pi}} 
\frac{1}{(1 + \delta_{K0})}
\nonumber\\
&
\times
\Bigg\{ 
\left< I' L K 0 | I K\right> 
\left[ 1\ +\ (-)^{(I + I')} r_I r_{I'} \right]
+ \ \delta_{K0}
\left< I' L 0 0 | I 0\right> 
\left[ (-)^{I} r_I\ +\ (-)^{I'} r_{I'} \right]
\Bigg\} 
\nonumber\\
&\hspace*{2.8cm}
=  
\sqrt{(2I'+1)} \frac{1}{(1 + \delta_{K0})}
\Bigg\{ \left< I' L K 0 | I K\right> 
\ + \ \delta_{K0} \left< I' L 0 0 | I 0\right> \Bigg\} ,
\label{Evrme}
\end{align}
as $(-)^I\ r_I = +1$ in all cases as evident from  Eq.~(\ref{roteq}). 
Thus, 
\begin{equation}
< \Psi^I_{K} \left\| Y_{L} \right\| \Psi^{I'}_{N'} >
=  \sqrt{\frac{(2I'+1)(2L+1)}{4\pi} }\ 
\left< I' L K 0 | I K \right> .
\end{equation}
For the $N=Z$ nuclei considered, the intrinsic energy is taken to be zero,
and the strongly coupled states are taken to be the $K=0$ ground state.
The structure factors are then simply the reduced matrix elements of three 
spherical harmonics.


\section{Using a vibration model of the nucleus}
\label{app3}

The surface of a liquid drop of incompressible fluid that can be slightly 
deformed 
is represented as
\begin{equation}
R(\theta \phi)  = R_0 \left[ \alpha_{0 0}^\star +
\sum_{\lambda \mu} \alpha_{\lambda \mu}^\star
Y_{\lambda \mu}(\theta \phi) \right] \ ,
\end{equation}
which, though similar to Eq.~(\ref{Rrotor}), has important differences.  As the 
radius  must be a real quantity, the coefficients must satisfy the spherical 
harmonic identity,
$\alpha_{\lambda \mu}^\star \equiv (-)^\mu \alpha_{\lambda -\mu}$.
The center of mass is defined by 
\begin{displaymath}
M{\bf R} = \sum_i m_i {\bf r_i} = \int \rho_m {\bf R} d{\bf r},
\end{displaymath}
where $\rho_m$ is the mass density assumed to be  uniform.
Considering the $z$-component of the center of mass coordinate, 
$Z = r \cos(\theta)$, which must be zero in the center of mass frame, we find
\begin{align}
MZ =& \sqrt{\frac{4\pi}{3}}\ \rho_m \int Y_{1 0}^\star(\Omega) r^2 dr
d\Omega
\nonumber\\ 
=& \sqrt{\frac{4\pi}{3}}\ \rho_m\ \frac{1}{4} R_0^4
\int  Y_{1 0}^\star(\Omega) 
\left[ \alpha_{0 0}^\star +
\sum_{\lambda \mu} \alpha_{\lambda \mu}^\star
Y_{\lambda \mu}(\theta \phi) \right]^3 d\Omega
\nonumber\\
\simeq& 
\sqrt{\frac{4\pi}{3}}\ \rho_m\ R_0^4 \ 
\left[ \alpha_{0 0}^\star\right]^3
\ \alpha_{1 0}^\star\hspace*{3.0cm}
{\rm(to\ first\ order)} .
\end{align}
As $\alpha_{0 0}$ is of order unity, for $Z$ to be zero, 
$\alpha_{1 0}^\star$ must vanish.  Hence, with a single fluid model, 
there can be no dipole ($\lambda = 1$) component in the expansion
of the surface.
Finally, as it is assumed that the drop is of incompressible matter,
the volume should remain constant.
This gives the constraint, 
\begin{align}
1 =& \frac{3}{4\pi R_0^3} \ \int\ r^2 dr\ d\Omega
\ =\ \frac{1}{4\pi}\int \left[ \alpha_{0 0}^\star +
\sum_{\lambda > 1, \mu} \alpha_{\lambda \mu}^\star
Y_{\lambda \mu}(\theta \phi) \right]^3\ d\Omega
\nonumber\\
=& \left[ \alpha_{0 0}^\star\right]^3
+ \frac{3}{4\pi}\ \alpha_{0 0}^\star\ \sum_{\lambda > 1, \mu} 
\left|\alpha_{\lambda \mu}\right|^2 + \cdots\cdots
\end{align}
Thus $\alpha_{0 0} = 1$ with correction only at second and higher orders
so that we take,
\begin{equation}
R(\theta \phi)  = R_0 \left[ 1 +
\sum_{\lambda > 1, \mu} \alpha_{\lambda \mu}^\star
Y_{\lambda \mu}(\theta \phi) \right] \ .
\label{Surface}
\end{equation}
With this specification of the nuclear surface, expansion to second 
order in deformation of any function gives,
\begin{align}
f(r) = f_0(r) &- R_0 \frac{df_0(r)}{dr}
\sum_{\lambda \mu} \alpha_{\lambda \mu}^\star Y_{\lambda \mu}(\theta \phi) 
\nonumber\\
&
+ \frac{1}{2} R_0^2 \ \frac{d^2f_0(r)}{dr^2}
\sum_{l_1 m_1 l_2 m_2} \alpha_{l_1 m_1}^\star \alpha_{l_2 m_2}^\star
Y_{l_1 m_1}(\theta \phi) Y_{l_2 m_2}(\theta \phi) .
\label{Vib-fst}
\end{align}
Therein, and in all that follows, it is presumed that summation of the
expansion labels of the generalised coordinates, and subsequently of the
angular momentum quantum numbers of the phonon creation/anihilation operators 
derived from them, exclude dipole terms, i.e. $\lambda > 1$.

The product of two generalised coordinates that satisfy the spherical harmonic 
condition, then can be written as,
\begin{align}
\alpha_{l_1 m_1}^\star \alpha_{l_2 m_2}^\star
=& \sum_{\nu_1 \nu_2}  \delta_{m_1 \nu_1} \delta_{m_2 \nu_2}
\alpha_{l_1 \nu_1}^\star \alpha_{l_2 \nu_2}^\star
\nonumber\\
=& \sum_{\lambda \mu}\ 
\left< l_1 l_2 m_1 m_2 | \lambda \mu \right>
\left[ \sum_{\nu_1 \nu_2} 
\left< l_1 l_2 \nu_1 \nu_2 | \lambda \mu \right>
\ \alpha_{l_1 \nu_1}^\star \alpha_{l_2 \nu_2}^\star\right]
\nonumber\\
=& \sum_{\lambda \mu}
\left< l_1 l_2 m_1 m_2 | \lambda \mu \right>\ 
\left[\alpha_{l_1}^\star \otimes \alpha_{l_2}^\star \right]_{\lambda
\mu}\ .
\label{contract-aa}
\end{align}
This form is convenient since 
$\left[\alpha_{l_1}^\star \otimes \alpha_{l_2}^\star \right]_{\lambda \mu}$ 
is a component of an irreducible tensor so that
the second order term in Eq.~(\ref{Vib-fst}) can be written as
\begin{align}
T_2 &= 
\frac{1}{2}
R_0^2\ \frac{d^2f_0(r)}{dr^2}\ 
\sum_{l_1 m_1 l_2 m_2 \lambda \mu K}
\left< l_1 l_2 m_1 m_2 | \lambda \mu \right>\
\left[\alpha_{l_1}^\star \otimes \alpha_{l_2}^\star \right]_{\lambda
\mu}
\nonumber\\ 
&\hspace*{1.0cm}\times\sqrt{\frac{(2l_1+1)(2l_2+1)}{4\pi(2K+1)}}
\ \left< l_1 l_2 0 0 | K 0 \right>
\ \left< l_1 l_2 m_1 m_2 | K M_K \right>
\ Y_{KM_K}(\Omega)\ ,
\end{align}
and which on using the orthogonality of Clebsch-Gordan coefficients
reduces to
\begin{equation}
T_2 = \frac{1}{2} R_0^2 
\ \frac{d^2f_0(r)}{dr^2}\ 
\sum_{\lambda}
\sqrt{\frac{(2l_1+1)(2l_2+1)}{4\pi(2\lambda + 1)}}
\ \left< l_1 l_2 0 0 | \lambda 0 \right>
\ \left[\alpha_{l_1}^\star \otimes \alpha_{l_2}^\star \right]_{\lambda}
{\bf \cdot  Y}_{\lambda}(\Omega)\ ,
\label{T2-contract}
\end{equation}
since the generalised coefficients must satisfy the spherical
harmonic condition.

Thus, matrix elements of the type,
\begin{align}
&\left[ f(r) \right]_{c c'} = 
\left[ f_0(r) \right]_{c c'}
- R_0 \frac{df_0(r)}{dr} \left[
\sum_\lambda \left[\alpha_\lambda^\star {\bf \cdot Y}_\lambda(\Omega)\right] 
\right]_{c c'}
\nonumber\\
&\; \; + {\frac{1}{2}} R_0^2 \ \frac{d^2f_0(r)}{dr^2} 
\left[ \sum_{l_1, l_2, \lambda} 
\sqrt{\frac{(2l_1+1)(2l_2+1)}{4\pi (2\lambda+1)}} 
\left< l_1 l_2 0 0 | \lambda 0 \right>
\left[\alpha_{l_1}^\star \otimes \alpha_{l_2}^\star\right]_\lambda
{\bf \cdot Y}_\lambda(\Omega) \right]_{c c'}
\label{frexpand}
\end{align}
are found.  The $\alpha_{\ell_i, m_{\ell_i}}$ are generalised 
(target) coordinates that
are quantised to be a combination of phonon creation and annihilation 
operators, i.e.
\begin{equation}
\alpha_{\lambda \mu} \Rightarrow \frac{1}{\sqrt{(2\lambda + 1)}}\ 
\beta_\lambda\ 
\left[b_{\lambda \mu} + (-)^\mu b^\dagger_{\lambda -\mu} \right]\ ,
\end{equation}
where $\beta_\lambda$ is the distortion parameter in this model.
Note that this specification differs in form and scale from
the development with rotation models.
A quantal phonon of vibration is created/annihilated
by the action of $b^\dagger_{LM}/b_{LM}$ on any initial state.
Thus, unlike the simpler rotation model, in this case we need to
specify expectation values of one and two phonon operators connecting
states described appropriately.  
They are considered later.

Later it is convenient to use the generalised forms,
\begin{align}
Q_{\lambda \mu}^{(1)} &= \alpha_{\lambda \mu}
\nonumber\\
Q_{\lambda \mu}^{(2)} &= \sum_{l_1 l_2}
\sqrt{\frac{(2l_1+1)(2l_2+1)}{4\pi (2\lambda+1)}} 
\left< l_1 l_2 0 0 | \lambda 0 \right>
\left[\alpha_{l_1}^\star \otimes \alpha_{l_2}^\star
\right]_{\lambda \mu} .
\label{vibops}
\end{align}

Using the basic interaction potential form given in 
Eqs.~(\ref{ch-state}-\ref{Eqn4}), the MCAS 
interaction matrix for the vibration model is
\begin{align}
\left\{ V\right\}_{c c'} =& 
\left[ \bigg( V_0 + V_{\ell \ell}\ \ell(\ell + 1)
\bigg) f_0(r)
 + \frac{1}{2}  V_{\ell I}\bigg(  
J(J+1) - \ell(\ell +1) - I(I+1)
\bigg) g_0(r) \right] \delta_{c c'}
\nonumber\\
&- R_0 \left[ \frac{df_0(r)}{dr} \bigg( V_0  +
\frac{1}{2} V_{\ell \ell} \left[
\ell' (\ell' + 1) + \ell (\ell + 1)\right]
\bigg) \right.
\nonumber\\
&\hspace*{0.5cm} \left.
+ \frac{1}{4} V_{\ell I} \frac{dg_0(r)}{dr} \bigg( 2 J(J + 1)
- \ell' (\ell' + 1) - \ell (\ell + 1)
- I' (I' + 1) - I (I + 1) \bigg) \right]
\nonumber\\
&\hspace*{10cm} \times
\sum_L \left[\alpha^\star_L {\bf \cdot Y}_L \right]_{c c'}
\nonumber\\
&+
\frac{1}{2} R_0^2\left[
\frac{d^2f_0(r)}{dr^2} \bigg( 
V_0 + 
\frac{1}{2} V_{\ell \ell} \left[
\ell' (\ell' + 1) + \ell (\ell + 1)
\right]  \bigg) \right.
\nonumber\\
&\hspace*{0.5cm} \left.
+ \frac{1}{4} V_{\ell I} \frac{d^2g_0(r)}{dr^2} \bigg( 
2 J(J + 1) - \ell' (\ell' + 1) - \ell (\ell + 1) - I' (I' + 1) - I (I + 1)
\bigg) \right]
\nonumber\\
&\hspace*{6.0cm}\times
\sum_\lambda \left[ \left\{\sum_{l_1 l_2} {\cal Q}_\lambda^{(2)}(l_1, l_2)
\right\} {\bf \cdot Y}_\lambda \right]_{c c'} ,
\label{Vcc-vib}
\end{align}
where
\begin{equation}
{\cal Q}^{(2)}_\lambda(l_1 l_2) =
\sqrt{\frac{(2l_1+1)(2l_2+1)}{4\pi (2\lambda+1)}}
\left< l_1 l_2 0 0 | \lambda 0\right>
\left[ \alpha_{l_1}^\star \otimes \alpha_{l_2}^\star \right]_\lambda .
\label{sec-calQ}
\end{equation}

The first and second order terms require development as matrix elements of 
nuclear phonon operators.  With scalar operators, 
$\left[ {\bf {\cal X}_L \cdot Y}_L(\Omega) \right]_{0,0}$,
using the identity, (Eq.~(5.13) in  ~\cite{Br68}), suitably 
adjusted to Edmond's form for the Wigner-Eckart theorem, 
\begin{align}
&\left< c \left| \left[ {\bf {\cal X}_L \cdot Y}_L(\Omega) \right]_{0,0}
\right| c' \right> =
\frac{1}{\sqrt{(2J + 1)}}\ 
\left< (\ell I)J \left\| \left[ {\bf {\cal X}_L \cdot Y}_L(\Omega) \right]_0
\right\| (\ell' I')J \right>
\nonumber\\
&\hspace*{3.4cm}= (-)^{(\ell' + I + J)} 
\left\{
\begin{array}{ccc}
\ell & \ell' & L\\
I' & I & J
\end{array}
\right\}
\left< \ell \left\| Y_L \right\| \ell'\right>
\left< I \left\| {\cal X}_L \right\| I'\right>
\nonumber\\
&\hspace*{0.4cm}=
(-)^{(\ell' + I + J)} 
\left\{
\begin{array}{ccc}
\ell & \ell' &  L\\
I' & I & J
\end{array}
\right\}
\sqrt{\frac{(2L+1)(2\ell'+1)}{4\pi}}
\left< \ell' L 0 0 | \ell 0 \right> 
\left< I \left\| {\cal X}_L \right\| I'\right> ,
\label{me-vibmod}
\end{align}
since
\begin{equation}
\left< \ell \left\| Y_L \right\| \ell' \right> = 
\sqrt{\frac{(2L+1)(2\ell'+1)}{4\pi}}
\left< \ell' L 0 0 | \ell 0\right> .
\end{equation}

With the above, Eq.~(\ref{Vcc-vib}) expands to
\begin{align}
\left\{ V\right\}_{c c'} =& 
\left[ \bigg( V_0 + V_{\ell \ell} \ell(\ell + 1) \bigg) f_0(r)
+ \frac{1}{2} V_{\ell I} \bigg( J(J+1) 
- \ell(\ell +1) - I(I+1)  \bigg) g_0(r) \right] \delta_{c c'}
\nonumber\\
&- R_0 \left[ \frac{df_0(r)}{dr} \bigg( V_0  + 
\frac{1}{2} V_{\ell \ell} \left[ \ell' (\ell' + 1) + \ell (\ell + 1)
\right] \bigg) \right.
\nonumber\\
&\hspace*{1.0cm} \left.
+ \frac{1}{4} V_{\ell I} \frac{dg_0(r)}{dr} \left[ 2 J(J + 1)
- \ell' (\ell' + 1) - \ell (\ell + 1) - I' (I' + 1) - I (I + 1)
\right] \right]
\nonumber\\
&\hspace*{3.0cm}\times
(-)^{(\ell' + I + J)}
\sum_L 
\left\{
\begin{array}{ccc}
\ell & \ell' & L \\
I' & I & J
\end{array}
\right\}
\sqrt{\frac{(2L+1)(2\ell'+1)}{4\pi}}
\nonumber\\
&\hspace*{8.5cm}
\times \left< \ell' L 0 0 | \ell 0\right>
\left< I \left\| \alpha_L^\star \right\| I'\right>
\nonumber\\
&+
\frac{1}{8\pi} R_0^2\ \left[
\frac{d^2f_0(r)}{dr^2}  \bigg(
V_0  + 
\frac{1}{2} V_{\ell \ell} \left[
\ell' (\ell' + 1) + \ell (\ell + 1)
\right] \bigg) \right.
\nonumber\\
&\hspace*{0.5cm} \left.
+ \frac{1}{4} V_{\ell I} \frac{d^2g_0(r)}{dr^2}  
\bigg( 2 J(J + 1)
- \ell' (\ell' + 1) - \ell (\ell + 1)
- I' (I' + 1) - I (I + 1)
\bigg) \right]
\nonumber\\
&\hspace*{2.0cm}\times
\ (-1)^{(\ell' + I + J)}
\sum_\lambda 
\sqrt{4\pi (2\ell' + 1)(2\lambda+1)}\ 
\left< \ell' \lambda 0 0 | \ell 0 \right>
\nonumber\\
&\hspace*{7.0cm}\times
\left\{
\begin{array}{ccc}
\ell & \ell' & \lambda \\
I' & I & J
\end{array}
\right\}
\left< I \left\| Q_\lambda^{(2)} \right\| I'\right> . 
\label{Vib-Vcc}
\end{align}


\bibliography{mcas-Alpha-A}

\end{document}